\documentclass[aps,pre,superscriptaddress,nofootinbib]{revtex4}

\usepackage{natbib,amsmath,amssymb,graphicx,subfigure,multirow,hyperref,color,lscape}

\long\def\symbolfootnote[#1]#2{\begingroup%
\def\thefootnote{\fnsymbol{footnote}}\footnote[#1]{#2}\endgroup}

\begin{document}

% v1 DJF 20/03/09
% v2 DJF 16/04/09
% 03, MAP, 17/04/09
% v4 DJF

\title{Dynamical Clustering of Exchange Rates}

\author{Daniel J. Fenn}
\email[To whom correspondence should be addressed: ]{fenn@maths.ox.ac.uk}
\affiliation{Mathematical and Computational Finance Group, Mathematical Institute, University of Oxford, Oxford OX1 3LB, UK}
\affiliation{CABDyN Complexity Centre, University of Oxford, Oxford OX1 1HP, UK}
\author{Mason A. Porter}
\affiliation{Oxford Centre for Industrial and Applied Mathematics, Mathematical Institute, University of Oxford, Oxford OX1 3LB, UK}
\affiliation{CABDyN Complexity Centre, University of Oxford, Oxford OX1 1HP, UK}
\author{Peter J. Mucha}
\affiliation{Carolina Center for Interdisciplinary Applied Mathematics, Department of Mathematics, University of North Carolina, Chapel Hill, NC 27599-3250, USA}
\affiliation{Institute for Advanced Materials, Nanoscience and Technology, University of North Carolina, Chapel Hill, NC 27599, USA}
\author{Mark McDonald}
\affiliation{FX Research and Trading Group, HSBC Bank, 8 Canada Square, London E14 5HQ, UK}
\author{Stacy Williams}
\affiliation{FX Research and Trading Group, HSBC Bank, 8 Canada Square, London E14 5HQ, UK}
\author{Neil F. Johnson}
\affiliation{Physics Department, University of Miami, Coral Gables, Florida 33146, USA}
\affiliation{CABDyN Complexity Centre, University of Oxford, Oxford OX1 1HP, UK}
\author{Nick S. Jones}
\affiliation{Physics Department, Clarendon Laboratory, University of Oxford, Oxford OX1 3PU, UK}
\affiliation{Oxford Centre for Integrative Systems Biology, Oxford OX1 3QU, UK}
\affiliation{CABDyN Complexity Centre, University of Oxford, Oxford OX1 1HP, UK}

\begin{abstract}
We use techniques from network science to study correlations in the foreign exchange (FX) market over the period 1991--2008. We consider an FX market network in which each node represents an exchange rate and each weighted edge represents a time-dependent correlation between the rates. To provide insights into the clustering of the exchange rate time series, we investigate dynamic communities in the network.  We show that there is a relationship between an exchange rate's functional role within the market and its position within its community and use a node-centric community analysis to track the time dynamics of this role. This reveals which exchange rates dominate the market at particular times and also identifies exchange rates that experienced significant changes in market role. We also use the community dynamics to uncover major structural changes that occurred in the FX market. Our techniques are general and will be similarly useful for investigating correlations in other markets.
\end{abstract}

\maketitle

%%%%%%%%%%%%%%%%%%%%%%%%%%%%%%%%%%%%%%%%%%%%

\section{Introduction}

Complex systems are composed of many interacting elements and can exhibit numerous forms of ``emergent" collective dynamics without the need for any external organizing principle \cite{BOCC_BOOK_2003}.  Such dynamics typically cannot be explained by studying the constituent parts in isolation, so a complex system must be analyzed as a whole.  Networks (i.e., graphs) provide a tractable framework for the quantitative analysis of many complex systems by distilling them to their key elements \cite{NEW_SIAM_2003,AMARAL_EPJB_2004,ALB_RMP_2002,CALD_BOOK_2007}. In such a representation, the elements of the system are represented as the network's nodes and the important interactions between them as links that connect the nodes.

Financial markets exhibit many of the key properties that characterize complex systems: They are composed of many heterogeneous components that interact with each other and their environment non-linearly in the presence of feedback \cite{AMARAL_EPJB_2004,MAN_ECON_2000}. An investigation of a financial market can then be formulated as a network problem.  Indeed, a wide range of financial assets have been investigated using network techniques, including equities \cite{MAN_EPJ_1999,MAN_ECON_2000,JPO_PRE_2003}, currencies \cite{MM_PRE_2005,MM_PRE_2008}, commodities \cite{SIEC_PHYSA_2009}, bonds \cite{BERN_PHYSA_2002}, and interest rates \cite{MATTEO_PHYSA_2004}. Network analyses have the potential to provide new insights into financial data and the structure of markets, which may in turn lead to the development of better market models.
% and any improvements in the understanding of market data enables one to develop better market models.

In the most common network description of a market, each node represents an asset, and each weighted link is a function (the same function for all links) of the pairwise temporal correlations between the two assets it connects \cite{MAN_ECON_2000}. In typical financial networks of $N$ assets, one can calculate a correlation coefficient between each pair of assets, so the network contains $\frac{1}{2}N(N-1)$ links.  Thorough, simultaneous investigation of the interactions is therefore difficult for even moderate $N$, so attaining an understanding of the market system necessitates some form of simplification.

The most prevalent method for reducing the complexity of a financial network is to construct a minimum spanning tree (MST) \cite{MAN_EPJ_1999,MAN_ECON_2000,JPO_PRE_2003,JPO_EPJ_2004,BOU_BOOK_2003}. The MST is generated using a hierarchical clustering algorithm \cite{DUDA_BOOK_2001} and reduces the network to $n-1$ of its most important microscopic interactions. This approach has resulted in many useful financial applications, including the construction of a visualization tool for portfolio optimization \cite{JPO_PRE_2003} and a means for identifying the effect of news and major events on market structure \cite{MM_PRE_2008}.  Nevertheless, the MST approach has a number of limitations which we discuss in Section \ref{sec::mst}.

An alternative simplification method is to coarse-grain the network and consider it at various mesoscopic scales. The properties of the market can then be understood by considering the dynamics of small groups of similar nodes. A widely-studied form of mesoscopic structure, known as a ``community'' \cite{NEW_EPJ_2004,NEWGIV_PRE_2004,DAN_JSM_2005,NEW_PRE_2006,RB_PRE_2006,AREN_ARX_2007,FORT_PNAS_2007,fortunato2009,MAP_SURVEY_2009}, is constructed from subsets of nodes that are more strongly connected to each other than they are to the rest of the network. Communities are of considerable interest to network scientists because they often correspond to behavioral or functional units \cite{MAP_PNAS_2005,TRAUD_2009,GUIM_NAT_2005,ADA_BIOINF_2006}, so their identification can lead to a better understanding of the function of dynamical processes (such as the spread of opinions and diseases) that operate on networks \cite{DAN_JSM_2005,fortunato2009,MAP_SURVEY_2009}.  From a financial perspective, communities correspond to groups of closely-related assets, so this treatment has the potential to suggest possible formulations for coarse-grained stochastic models of markets.

During the last decade, there has been an explosion of papers on networks with static connections between nodes, and research on dynamical systems on such networks has now also become ubiquitous \cite{NEW_SIAM_2003,CALD_BOOK_2007,book2008}. However, there has been much less research on networks that are themselves time-dependent \cite{VICSEK_Nat_2007,jp,multislice}, and a characterization of such networks is essential to fully understand dynamical processes on networks. One of the main reasons for the limited analysis of time-dependent networks is the difficulty of acquiring time-dependent data. Fortunately, financial markets are one of the most data-rich complex systems, providing a valuable source of accurate, high-frequency, time-series data. Financial data is therefore an important resource for developing tools and theories for describing time-evolving networks.

In the present work, we build on the results of our recent short paper \cite{FENN_2009} and investigate community dynamics in a time-evolving foreign exchange (FX) market network over three periods: 1991--1998 (before the introduction of the euro), 1999--2003 (following the introduction of the euro), and 2005--2008 (which includes the recent credit and liquidity crisis).\footnote{We do not have data for 2004.} The FX network possesses a fixed number of nodes and evolving link weights that are determined by time-varying pairwise correlations between time series associated with each node. Therefore, in contrast to some other studies of financial networks, we analyze the fully-connected network and do not remove links below some threshold \cite{FAR_NJP_2007}. Community detection in networks of this kind is equivalent to the problem of clustering multivariate time series \cite{TS_CLUSTERING}. We also track the communities from the perspective of individual nodes, which removes the undesirable requirement of determining which community at each time-step represents the descendant of a community at the previous time-step. Previous dynamic community studies have attempted to track entire communities \cite{VICSEK_Nat_2007,HOP_PNAS_2004} but (as discussed in Section~\ref{sub::tracking}) some of these approaches can lead to equivocal mappings following community splits and mergers.

We demonstrate that exchange rate community dynamics provides insights into the correlation structures within the FX market and uncovers the most important exchange rate interactions. We also show that large community reorganizations often accompany significant market events and that the details of such community adjustments can reveal the trading behavior that leads to these changes. We find that there is a relationship between an exchange rate's functional role within the market and its position within its community, and we identify exchange rates that experience significant changes in market role. Although we focus on the FX market, the techniques that we present are general and will be similarly insightful for other asset classes. We note that this paper extends the results described in \cite{FENN_2009}.

The remainder of this paper is organized as follows.  In Section \ref{sec::data}, we discuss the nature of the FX data and use it to derive a time-dependent network. We detect communities in the network in Section \ref{sec::detection} and discuss robust communities in Section \ref{sec::robust}.  We examine the properties of the communities in Section \ref{sec::dynamics}, compare the detected communities to minimum spanning trees in Section \ref{sec::mst}, and derive the roles of exchange rates within communities in Section \ref{sec::centroles}.  We relate the major changes in community structure over time to significant changes in the FX market in Section \ref{sec::majorchange} and investigate the changes in the community roles of exchange rates in Section \ref{sec::rolevis}. In Section~\ref{sec:heuristics}, we discuss the effects of using different heuristics to identify optimal partitions of the networks into communities. In Section \ref{sec::conclusions} we offer some conclusions.

%%%%%%%%%%%%%%%%%%%%%%%%%%%%%%%%%%%%%%%%%%%%

\section{Data} \label{sec::data}
The FX networks we construct have $n=110$ nodes, each of which represents an exchange rate of the form XXX/YYY (with XXX$\neq$YYY), where XXX, YYY$\in$\{AUD, CAD, CHF, GBP, DEM, JPY, NOK, NZD, SEK, USD, XAU\}\footnote{These symbols represent: AUD, Australian dollar; CAD, Canadian dollar; CHF, Swiss franc; EUR, euro; GBP, pounds sterling; JPY, Japanese yen; NOK, Norwegian krone; NZD, New Zealand dollar; SEK, Swedish krona; USD, U.S. dollar; XAU, gold. We include gold in the study because it has many similarities with a currency \cite{MM_PRE_2005}.} and we note that DEM$\rightarrow$EUR after 1998. An exchange rate XXX/YYY indicates the amount of currency YYY that can be received in exchange for one unit of XXX.\footnote{For each exchange rate both a bid and an ask price will be quoted. Bid/ask prices give the different prices at which one can buy/sell currency, with the ask price tending to be larger than the bid price. For example, the exchange rate between EUR and USD might be quoted as $1.4085$/$1.4086$. A trader then looking to convert USD into EUR might have to pay $1.4086$ USD for each EUR, whereas a trader looking to convert EUR to USD might receive only $1.4085$ USD per EUR.} Other authors have recently studied the FX market by constructing networks in which all nodes represent exchange rates with the same base currency, implying that each node can then be considered to represent a single currency \cite{GORSKI_EPJB_2008}. Exchange rate networks formed with reference to a single base currency are somewhat akin to ego-centered networks studied in the social networks literature \cite{WASSERMAN_BOOK_2008}. Ego-centered networks include links between a number of nodes that all have ties to an \textit{ego} which is the focal node of the network. However, this approach has two major problems for FX networks. First, it neglects a large number of exchange rates that can be formed from the set of currencies studied and consequently also ignores the interactions between these rates. Second, the network properties depend strongly on the choice of base currency and this currency is, in effect, excluded from the analysis. We therefore construct a network including all exchange rates that can be formed from the studied set of currencies.

The return of an exchange rate with price $p_i(t)$ at discrete time $t$ is defined by
\begin{equation}
	R_i(t)=\ln{\frac{p_i(t)}{p_i(t-1)}}\,.
\end{equation}
We take the price $p_i(t)$ as the mid-price of the bid and ask prices, so that
\begin{equation}
	p_i(t)=\dfrac{p_i^\textrm{bid}(t)+p_i^\textrm{ask}(t)}{2}\,.
\end{equation}
We use the last posted price within an hour to represent the price for the following hour. To calculate a return at time $t$, one needs to know the price at both $t$ and $t-1$. To minimize the possibility of a price not being posted in a given hour, we focus on the FX market's most liquid period: 07:00-18:00 U.~K. time. Nevertheless, there are still hours for which we do not have price data (this usually occurs as a result of problems with the data feed). One can calculate a return for hours with missing price data by assuming the last posted price or interpolating between prices at the previous and next time-step \cite{DAC_HFF_2001}. However, to ensure that all time-steps included in the study are ones at which a trade can actually be made, we take the stricter approach of omitting all returns for which one of the prices is not known.  In order to ensure that the time series of exchange rates are directly comparable, we consequently remove a return from all exchange rates if it is missing from any rate.
%if a return is missing for one of the exchange rates we also remove it for all of the other rates.

For the period 1991--2003, we derive each exchange rate XXX/YYY with XXX, YYY$\neq$USD from two USD rates. For example, we find the CAD/CHF price at each time-step by dividing the USD/CHF price by the USD/CAD price. For the period 2005--2008, we derive each exchange rate not included in the set \{AUD/USD, EUR/NOK, EUR/SEK, EUR/USD, GBP/USD, NZD/USD, USD/CAD, USD/CHF, USD/JPY, USD/XAU\} from pairs of exchange rates in this set. For example, we find the USD/NOK price at each time-step by dividing the EUR/NOK price by the EUR/USD price. Although this approach appears somewhat artificial, it matches the way in which many exchange rates are calculated in the actual FX market. For example, a bank customer wishing to convert CAD to NZD (or vice versa) will need to be quoted the CAD/NZD prices. Because this is not a standard conversion, the bank will not be able to quote a direct market price but will instead calculate a price using the more widely traded USD/NZD and USD/CAD exchange rates. Calculating the exchange rates in this way implies that there is some intrinsic structure inherent in the FX market.  However, as shown in \cite{MM_PRE_2005} and demonstrated further in Sections~\ref{sub::sigtest} and \ref{sub::properties} of this paper, this ``triangle effect'' does not dominate the results.

We determine the weights of the edges connecting pairs of nodes in the network using a function of the linear correlation coefficient $\rho$ between the return time series for the corresponding exchange rates. The correlation between the returns of exchange rates $R_i$ and $R_j$ over a time window of $T$ returns is given by
\begin{equation}
\rho(i,j)=\dfrac{\langle{R_iR_j}\rangle-\langle{R_i}\rangle\langle{R_j}\rangle}{\sigma_i\sigma_j}\,,
\end{equation}
where $\langle\cdot\rangle$ indicates a time-average over $T$ returns and $\sigma_i$ is the standard deviation of $R_i$ over $T$. We use the linear coefficient $\rho(i,j)$ to measure the correlation between pairs of exchange rates because of its simplicity, but one could use alternative measures that are capable of detecting more general dependencies \cite{SCHE_BOOK_2006}. Our methods can be applied using any choice for $\rho(i,j)$. The weighted adjacency matrix $\mathbf{A}$ representing the network then has components
\begin{equation}
	A_{ij} = \frac{1}{2}\Big{(}\rho(i,j)+1\Big{)}-\delta_{ij}\,,
\label{network}
\end{equation}
where the Kronecker delta $\delta_{ij}$ removes self-edges. The matrix elements $A_{ij}\in[0,1]$ quantify the similarity of each pair of exchange rates $i$ and $j$. For example, two exchange rates $i$ and $j$ whose return time series are perfectly correlated will be connected by a link of unit weight. (See Section \ref{sec::detection} for a discussion of alternative forms for $A_{ij}$.)

We exclude self-edges in order to deal with simple graphs. This approach was also taken in a previous study of a stock network derived from a correlation matrix \cite{HEIM_ARX_2008}. We note that if we include self-edges, the node compositions of the identified communities are identical if one makes a small parameter change in the community detection algorithm. We discuss the community detection algorithm and the effect of including self-edges in Sections \ref{sec::detection} and \ref{sec::dynamics}.

\begin{figure}
\includegraphics[width=0.85\linewidth]{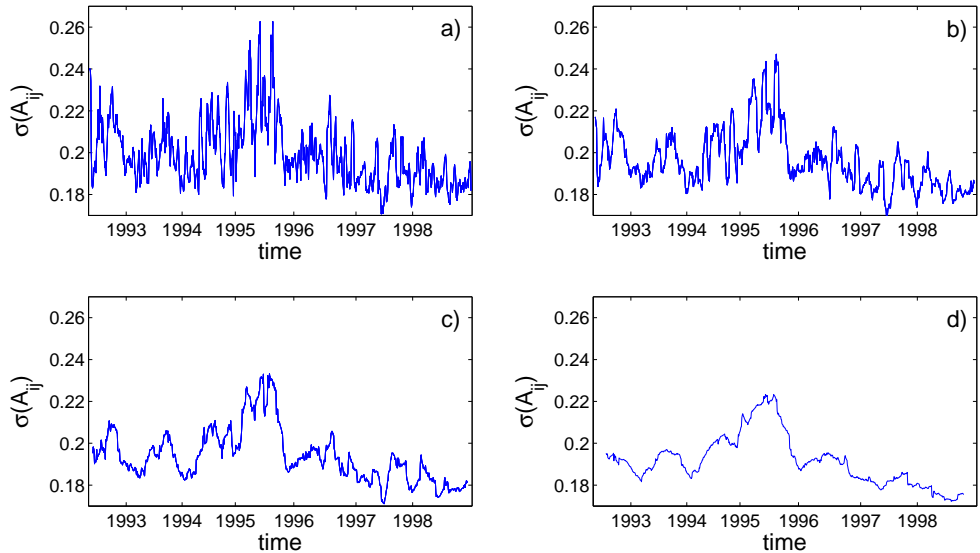}
\caption{\label{Teffect}(Color online) The standard deviation of the edge weights $A_{ij}$ as a function of time for the period 1991--1998. For each panel, $\Delta{t}=20$ (approximately 2 days), and (a) $T=100$ hours, (b) $T=200$ hours, (c) $T=400$ hours, and (d) $T=1200$ hours (approximately $0.5$, $1$, $2$, and $6$ months,  respectively).}
\end{figure}

\begin{figure}
\includegraphics[width=0.85\linewidth]{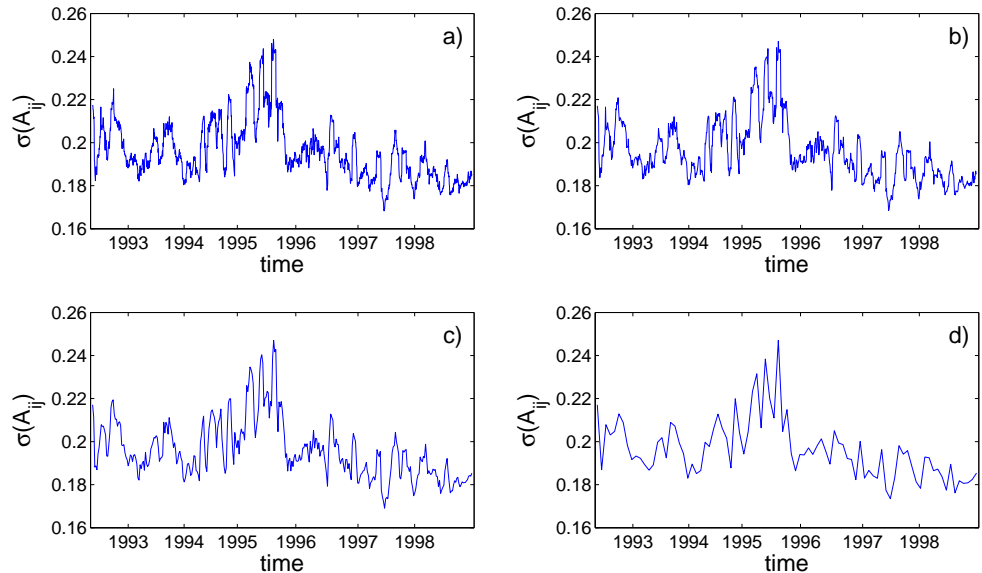}
\caption{\label{dteffect}(Color online) The standard deviation of the edge weights $A_{ij}$ as a function of time for the period 1991--1998. For each panel, $T=200$ hours, and (a) $\Delta{t}=10$, (b) $\Delta{t}=20$, (c) $\Delta{t}=50$, and (d) $\Delta{t}=200$ (approximately $1$ day, $2$ days, $5$ days, and $2$ weeks, respectively).}
\end{figure}

We create a longitudinal sequence of networks by consecutively displacing the time windows by $\Delta{t}=20$ hours (approximately 2 trading days) and fix $T=200$ hours (approximately 1 month of data). This choice of $T$, motivated in part by the example data in Fig.~\ref{Teffect}, represents a trade-off between over-smoothing for long time windows and overly-noisy correlation coefficients for small $T$ \cite{JPO_PRE_2003}. Figure \ref{dteffect} demonstrates that the choice of $\Delta{t}$ has a similar, but less pronounced, effect on the standard deviation of the edge weights and we again select a compromise value. The time windows we use to construct the networks overlap, so the single-time networks are not independent but rather form an evolving sequence through time.

%%%%%%%%%%%%%%%%%%%%%%%%%%%%%%%%%%%%%%%%%%%%

\section{Community detection} \label{sec::detection}

Communities consist of cohesive groups of nodes that are more strongly connected to each other than they are to the rest of the network and thus often represent functionally-important subnetworks \cite{girvan,DAN_JSM_2005,fortunato2009,MAP_PNAS_2005,TRAUD_2009,GUIM_NAT_2005,ADA_BIOINF_2006,MAP_SURVEY_2009}.  Most prior studies of financial networks find groups of closely-related assets using traditional hierarchical clustering techniques \cite{MAN_ECON_2000,JPO_PRE_2003,MM_PRE_2005} or by thresholding to create a binary network \cite{FAR_NJP_2007}. In this paper, we identify communities in high frequency, time-evolving, weighted networks using a technique based on the maximization of a quality function known as \textit{modularity} \cite{NEWGIV_PRE_2004}. To our knowledge, other papers with similar approaches have not examined longitudinal networks or have considered networks of equities rather than exchange rates \cite{HEIM_ARX_2008}. Dynamic communities have been investigated in biophysical data using methods based on modularity maximization \cite{SHALIZI_BOOK_2007}.  However, Ref.~\cite{SHALIZI_BOOK_2007} is concerned with the dynamics of functional communities that arise from coordinated behaviors taking place on the network rather on than the community dynamics of the underlying network. (In Section~\ref{sub::tracking}, we discuss other work that investigates community dynamics in non-financial data using alternative community detection techniques.)

The identification of communities using graph modularity is based on the idea that random networks are not expected to demonstrate community structure beyond small fluctuations. Modularity therefore identifies communities by finding subsets of nodes that are more strongly connected to each other than one would expect for a null model. Let $C$ be a partition of the $n$ nodes in $\mathbf{A}$ into mutually disjoint communities. The modularity $Q$ of the partition $C$ is given by
\begin{equation}
	Q(C)=\frac{1}{2m}\sum_{ij}(A_{ij}-P_{ij})\delta(c_i,c_j)\,, \label{modularity}
\end{equation}
where $c_i$ is the community of node $i$ and $P_{ij}$ denotes the expected weight of the link with which nodes $i$ and $j$ are connected in a null model. The quantity $m$ represents the total edge weight in the network and is given by $m=\frac{1}{2}\sum_{i}k_{i}$, where $k_i=\sum_jA_{ij}$ is the strength (weighted degree) of node $i$. Communities are identified by finding the partition $C$ that maximizes $Q$. The most popular choice of null model is the Newman-Girvan (NG) model \cite{NEWGIV_PRE_2004}
\begin{equation}
	P_{ij}=\frac{k_ik_j}{2m}\,, \label{config}
\end{equation}
which preserves the strength distribution of the network and is closely related to the configuration model \cite{MOLL_RSA_1995}. An alternative null model is a uniform model in which a fixed average edge weight occurs between nodes \cite{MAP_SURVEY_2009}.

%One can therefore identify network communities by finding the partition of the network that maximizes the modularity.
%The choice of null model is not entirely unconstrained because it is axiomatically the case that $Q=0$ when all of the nodes are placed in a %single group. One is then restricted to null models in which the expected edge weight is equal to the actual edge weight in the original network %\cite{NEW_PRE_2006}. The simplest null model satisfying this criterion is that of a random graph in which each edge has the same expected %weight. However, the strength distribution produced by this model is significantly different to the distribution observed for many real-world %networks. A more commonly used null model is the Newman and Girvan (NG) model \cite{NEWGIV_PRE_2004}, which preserves the strength distribution %of the network under study. The NG model is defined by

We construct FX networks by calculating a correlation coefficient between every pair of exchange rates, resulting in a weighted, fully-connected network.
 %(i.e., every node is connected to every other node). 
We include each exchange rate XXX/YYY and its inverse rate YYY/XXX in the network, because one cannot infer a priori whether a rate XXX/YYY will form a community with a rate WWW/ZZZ or its inverse ZZZ/WWW. However, the return of an exchange rate XXX/YYY is related to the return of its inverse YYY/XXX by $R_{\frac{XXX}{YYY}}=-R_{\frac{YYY}{XXX}}$.  This implies that the correlation coefficients between these rates and a rate WWW/ZZZ are related by $\rho(\frac{XXX}{YYY},\frac{WWW}{ZZZ})=-\rho(\frac{YYY}{XXX},\frac{WWW}{ZZZ})$.  Consequently, every node has the same strength
\begin{equation}
	k_i=\sum_jA_{ij}=\frac{1}{2}(n-2)\,,
\label{strength}
\end{equation}
so the probability of connection in the NG null model $P_{ij}=k_ik_j/2m$ is also constant and is given by
\begin{equation}
P_{ij}=\frac{n-2}{2n}.
\end{equation}
In the case of the FX network, the NG model and the uniform null model are thus equivalent.  However, the methods we present are general and can be applied to networks with non-uniform strength distributions. Additionally, every community has an equivalent inverse community. For example, if there is a community consisting of the three exchange rates XXX/YYY, XXX/WWW, and ZZZ/WWW in one half of the network, there must be an equivalent community formed of YYY/XXX, WWW/XXX, and WWW/ZZZ in the other half. The existence of an equivalent inverse community for each community means that at each time-step, the network is composed of two equivalent halves.  However, the exchange rates residing in each half change in time as the correlations evolve.

An important issue with using modularity as a quality function to identify communities is that modularity optimization has been shown to fail to find communities smaller than a scale that depends on the total size of the network and on the degree of interconnectedness between the network communities \cite{FORT_PNAS_2007}. However, many modularity-optimization techniques can easily be adapted to other quality functions, and several alternatives have been proposed that avoid the resolution limit by uncovering communities at multiple resolutions \cite{RB_PRE_2006,NEW_PRE_2006,AREN_ARX_2007,FOR_ARX_2008}.

In \cite{RB_PRE_2006}, Reichardt and Bornholdt proposed a multiresolution method in which the network $\mathbf{A}$ is represented as an infinite-range, $n$-state Potts spin glass in which each node is a spin, each edge is a pairwise interaction between spins, and each community is a spin state. The Hamiltonian of this system is given by
\begin{equation}
	\mathcal{H}(\gamma)=-\sum_{ij}J_{ij}\delta(c_i,c_j)\,, \label{POTTS}
\end{equation}
where $c_i$ is the state of spin $i$ and $J_{ij}$ is the interaction energy between spins $i$ and $j$. The coupling strength $J_{ij}$ is given by $J_{ij}=A_{ij}-{\gamma}P_{ij}$, where  $P_{ij}$ again denotes the expected weight of the link with which nodes $i$ and $j$ are connected in a null model and $\gamma$ is a resolution parameter. One can find communities by assigning each spin to a state and minimizing the interaction energy of these states given by Eq.~(\ref{POTTS}). Within this framework, community identification is equivalent to finding the ground state configuration of a spin glass.

Tuning $\gamma$ allows one to find communities at different resolutions. As $\gamma$ becomes larger, there is a greater incentive for nodes to belong to smaller communities. The Potts method therefore allows the investigation of communities below the resolution limit of modularity. One can write a scaled energy $Q_s$ in terms of the Hamiltonian in Eq.~(\ref{POTTS}) as
\begin{equation}
	Q_s = \frac{-\mathcal{H}(\gamma)}{2m}\,. \label{Q_POTTS}
\end{equation}
The modularity is then the scaled energy with $\gamma=1$. Community detection using modularity optimization is therefore a special case of the Potts method.

Recently, an alternative version of the Potts method has been proposed that is able to deal with both positive and negative links \cite{TRAAG_ARX_2008}. One can apply this technique to FX data using the correlation matrix $\mathbf{\rho}$ as the network adjacency matrix.  Using this approach and a uniform null model, we found the same robust communities (see Section~\ref{sec::robust} for a discussion of robust communities) as we identified using the Potts method and the adjacency matrix in Eq.~(\ref{network}). However, the Potts method for signed adjacency matrices did not identify the same robust communities when we employed the NG null model.

In this paper, we use the Potts method to detect communities of exchange rates in FX networks with adjacency matrices given by Eq.~(\ref{network}), and we employ the NG model of random link assignment $P_{ij}=k_ik_j/2m$ as the prior.\footnote{If we include self-edges in the network, the strength of each node increases by one. This, in turn, leads to a constant increase in the expected edge weight in the null model. For a network with self-edges, the expected edge weight is given by $P^{s}_{ij}=n/[2(n+2)]$, a shift by a constant value of $P_{ij}^s-P_{ij}= 2/[n(n+2)]\approx1.62\times{10^{-4}}$ relative to the network in which self-edges are excluded. Self-edges always occur within a community, so they will always contribute to the summation in Eq.~(\ref{POTTS}) irrespective of exactly how the nodes are partitioned into communities. This implies that self-edges play no role when determining the community partition that minimizes the interaction energy at a particular resolution.} The number of possible community partitions grows at least exponentially with the number of nodes \cite{NEW_PRE_2004_066133}, so it is typically impossible computationally to sample the energy space by exhaustively enumerating all partitions \cite{BRANDES_2006}. A number of different heuristic procedures have been proposed to balance the quality of the identified optimal partition with computational costs \cite{MAP_SURVEY_2009,DAN_JSM_2005,fortunato2009}. We minimize Eq.~(\ref{POTTS}) at each resolution using the greedy algorithm of Ref.~\cite{BLOND_ARX_2008}. We discuss the effect on our results of using different optimization heuristics in Section~\ref{sec:heuristics}.

%%%%%%%%%%%%%%%%%%%%%%%%%%%%

\section{Robust community partitions} \label{sec::robust}

In many networks, the same community structure persists across a range of resolutions \cite{RB_PRE_2006,AREN_ARX_2007,fortunato2009}. As one increases the resolution parameter in the Potts method one is providing an energy incentive for nodes to belong to smaller clusters. Community partitions that are robust across a range of resolutions are therefore significant because the communities do not break up despite an increasing incentive to do so. Communities in robust partitions have been found to correspond to the communities imposed by construction in simulated networks and to known groupings in real-world networks \cite{AREN_ARX_2007,fortunato2009}. This suggests that the communities in partitions that persist over a large range of resolutions potentially represent important substructures.

We compare community partitions using the normalized variation of information $\hat{V}$ \cite{MEIL_JMA_2006,TRAUD_2009}. The entropy of a partition $C$ of the $n$ nodes in $\mathbf{A}$ into $K$ communities $c^k$ ($k\in\{1,\cdots,K\}$) is
\begin{equation}
	S(C)=-\sum_{k=1}^{K}p(k)\log{p(k)}\,,
\label{entropy}
\end{equation}
where $p(k)={\vert{c^k}\vert}/{n}$ is the probability that a randomly-selected node belongs to community $k$ and $\vert{c^k}\vert$ is the size (set cardinality) of communities.\footnote{Note that the quantity $c^k$ represents the set of communities indexed by $k$ but that $c_i$ is the set of nodes in the same community as node $i$.} For a partition $C$, the entropy therefore indicates the uncertainty in the community membership of a randomly-chosen node. Given a second partition $C'$ of the $n$ nodes into $K'$ communities, the mutual information $I(C,C')$ is given by
\begin{equation}
	I(C,C')=\sum_{k=1}^{K}\sum_{k'=1}^{K'}p(k,k')\log\frac{p(k,k')}{p(k)p(k')}\,,
\end{equation}
where $p(k,k')=\vert{c^{k}\cap{c^{k'}}\vert}/{n}$. The mutual information is the amount by which knowledge of a node's community in $C$ reduces the uncertainty about its community membership in $C'$ (averaged over all nodes). The normalized variation of information $\hat{V}$ between $C$ and $C'$ is then given by
\begin{equation}
	\hat{V}(C,C')=\frac{S(C)+S(C')-2I(C,C')}{\log{n}}\,. \label{normvi}
\end{equation}
The factor $\log{n}$ normalizes $\hat{V}(C,C')$ to the interval $[0,1]$, with $0$ indicating identical partitions and $1$ indicating that all nodes are in individual communities in one partition and in a single community in the other.  We will use Eq.~(\ref{normvi}) to compare partitions in networks with the same number of nodes and remark that one should not normalize by $\log{n}$ when comparing the variation of information in data sets with different sizes \cite{MEIL_JMA_2006}.

The variation of information is a desirable measure for quantifying the difference between partitions of a network because it is a metric on the space of community assignments and satisfies the triangle inequality. Therefore, if two partitions are close to a third partition, they cannot differ too much from each other. It is also a local measure, so the contribution to $\hat{V}(C,C')$ from changes in a single community does not depend on how the rest of the nodes are clustered \cite{KARR_ARX_2007,MEIL_JMA_2006}.

\begin{figure}
\includegraphics[width=0.85\linewidth]{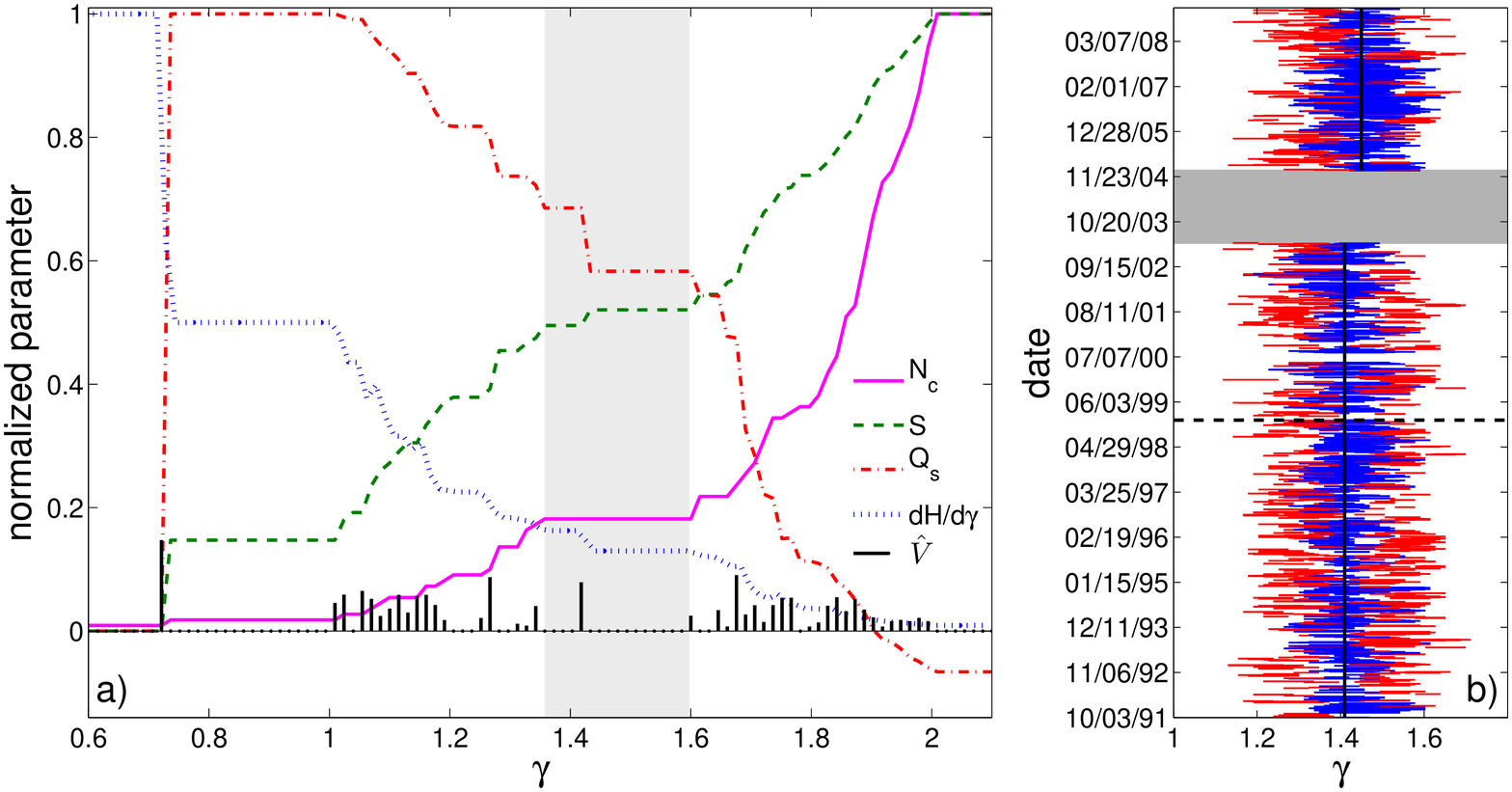}
\caption{\label{plateg}(Color online) (a) The quantities $N_c$, $S$, $Q$, and $d\mathcal{H}/d\gamma$ (defined in the text), normalized by their maximum values, versus the resolution parameter $\gamma$ for a single time window beginning on 03/17/1992. The shaded gray area highlights the main plateau. The bottom curve gives the normalized variation of information between partitions at resolutions separated by $\Delta\gamma=0.015$. (b) The position of the main plateau at each time-step for the full period 1991--2008. Main plateaus (blue) containing the fixed resolution (set at $\gamma=1.41$ for 1991--2003 and to $\gamma=1.45$ for 2005--2008) and (red) not containing the fixed resolution. The gray block corresponds to 2004, for which we do not have data.}
\end{figure}

One can identify robust communities by detecting communities at multiple resolutions and calculating $\hat{V}(C,C')$ between the community partitions for consecutive resolutions. Robust communities are revealed by intervals in which there are few spikes in $\hat{V}(C,C')$. In Fig.~\ref{plateg}(a) we show $\hat{V}(C,C')$ between community partitions detected at 100 resolutions in the interval $\gamma\in[0.6,2.1]$ separated by $\Delta\gamma=0.015$. We focus on this interval in this example because at $\gamma=0.6$ all of the nodes are assigned to the same community and at $\gamma=2.1$ all of the nodes are assigned to singleton communities. One can also identify robust communities by examining summary statistics that describe the community structure as a function of the resolution parameter. We consider the number of communities $N_c$, the modularity $Q$ (see Eq.~(\ref{modularity})), the entropy $S$ (see Eq.~(\ref{entropy})), and the rate of change of the energy with resolution $d\mathcal{H}/d\gamma$ (see Eq.~(\ref{POTTS} for the definition of $\mathcal{H}$). Robust communities correspond to plateaus (constant values) in curves of any of these quantities as a function of the resolution parameter. In Fig.~\ref{plateg}(a), we plot curves for each of the summary statistics as a function of $\gamma$.

Figure \ref{plateg}(a) contains four principle plateaus, corresponding to partitions of the network into $N_c=1$, $2$, $20$, and $110$ communities. The first and last plateaus, respectively, represent all nodes in a single community and all nodes in individual communities. The second plateau represents one community of exchange rates and a corresponding community of inverse rates. The $N_c=20$ plateau occurs over the interval $\gamma\in[1.34,1.57]$, in which there is a single plateau in the $N_c$ plot and a few smaller plateaus in each of the other plots. In contrast to the other plateaus, this one was not expected, so the robust communities over this interval can potentially provide new insights into the correlation structure of the FX market. Although the community configuration over this interval does not have maximal $Q$ (i.e., it is not the community configuration corresponding to the maximum value of the traditional modularity, which is the scaled energy with $\gamma=1$.), it provides an appropriate resolution at which to investigate community dynamics due to its resolution robustness and the financially-interesting features of the detected communities. For the remainder of this paper, we will refer to this plateau as the ``main'' plateau.

%This property also provides a good test for the detection algorithm because one always expects to find these equivalent communities. The greedy %search algorithm finds community configurations that are close to the optimal configuration, but that do not always correspond exactly to it. %Often the algorithm doesn't find a configuration consisting of a community of exchange rates and an equivalent community of inverse rates, but %the two halves differ in the community assignment of one or two nodes. The configuration with these nodes reassigned such that the communities %in the two halves of the network are exactly equivalent is lower in energy, and so closer to the ground state, than the configuration originally %found by the algorithm.

%%%%%%%%%%%%%%%%%%%%%%%%%%%%%%%%%%%%%%%%%%%%

\section{Dynamic community detection} \label{sec::dynamics}

\subsection{Choosing a resolution} \label{sub::resolution}

\begin{figure}
\includegraphics[width=0.85\linewidth]{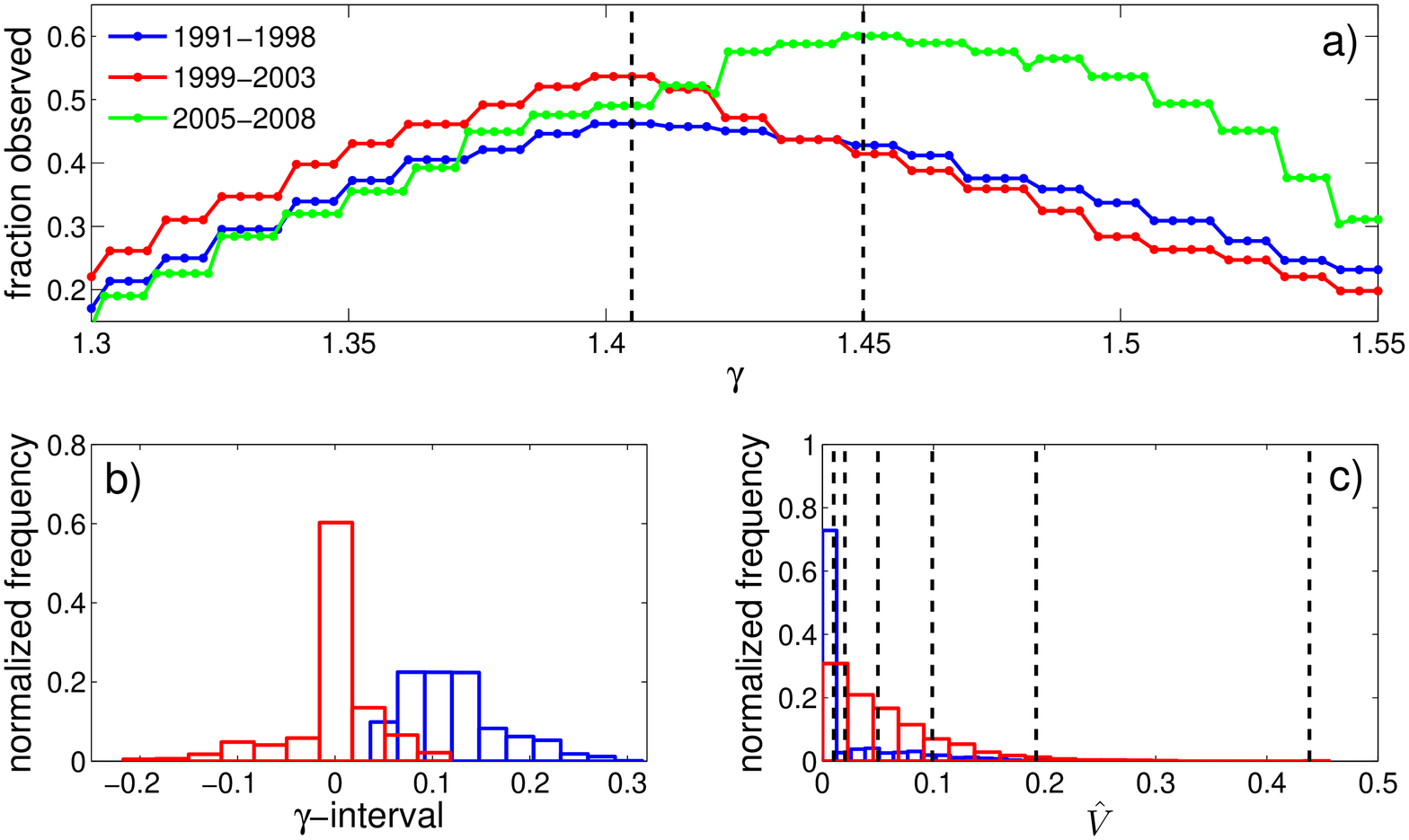}
\caption{\label{platStats}(Color online)
(a) Observed fraction of time-steps that the resolution $\gamma$ lies on the main plateau. The vertical lines indicate $\gamma=1.41$, which lies in the highest number of main plateaus for the period 1991--2003, and $\gamma=1.45$, which lies in the highest number of main plateaus for 2005--2008. These are the resolutions at which we investigate the community dynamics over the two periods. For the full period 1991--2008, we show in panel (b) the normalized sampled distribution of the main plateau width (blue) and the normalized sampled distribution of the $\gamma$-distance between the main plateau and the fixed resolution (red). The distance is exactly zero for 53\% of the time-steps.  Again for 1991-2008, we show in panel (c) the normalized variation of information distribution between the community configuration at the fixed resolution and the configuration corresponding to the main plateau (blue) and the normalized variation of information distribution between consecutive time-steps (red). The value of $\hat{V}$ is exactly zero for 64\% of the time-steps. The vertical lines give the mean $\hat{V}$ when (left to right) $1$, $2$, $5$, $10$, $20$, and $50$ nodes are randomly reassigned to different communities (averaged over 100 reassignments for each time-step).}
\end{figure}

To investigate the community dynamics, we first choose a resolution parameter at which to detect communities at each time-step. One approach is to always select a resolution $\gamma$ in the main plateau. As shown in Figs.~\ref{plateg}(b) and \ref{platStats}(a), this plateau occurs over different $\gamma$ intervals at different time-steps and has different widths. These intervals need not share common resolution values, so this method seems inappropriate because one would then be comparing communities obtained from many different resolutions. Therefore, we fix the resolution at the value that occurs within the largest number of main plateaus. As shown in Fig.~\ref{platStats}(a), this corresponds to $\gamma=1.41$ for the period 1991--2003 and $\gamma=1.45$ for the period 2005--2008.\footnote{In order to find equivalent communities in the network in which self-edges are included, it is necessary to decrease the resolution parameter to compensate for the increase in the constant expected edge weight in the null model. If we identify communities in the network in which self-edges are excluded using the resolution parameter $\gamma$, then we find identical communities in the corresponding network with self-edges using a resolution parameter $\gamma^{s}=\gamma{p_{ij}}/{p^{s}_{ij}}=\gamma(n+2)(n-2)/n^2$. For example, if we identify communities in the network without self-edges using a resolution of $\gamma=1.4500$, then we identify equivalent communities in the network with self-edges with a resolution parameter of $\gamma_s=1.4495$.}
%\footnote{These periods are treated as separate examples.}
%\footnote{It should be noted that the period 2005--2008 is investigated at a different resolution to the other two periods, which seems contrary to our argument that it is inappropriate to compare communities obtained from different resolutions. We will, however, only be making very broad aggregated comparisons between the three periods and will not be comparing individual time-steps.}
%{\bf map: let's only make a comment here if we are asked; the fact that we don't have data for 2004 suggests that we should treat these as different examples anyway}

In order to demonstrate the validity of this technique, we show in Fig.~\ref{platStats}(b) the distribution of the $\gamma$ distance from the fixed resolution to the main plateau and in Fig.~\ref{platStats}(c) the distribution of the normalized variation of information between the community configuration obtained at the fixed resolution and that corresponding to the main plateau. Both distributions are strongly peaked at zero. The fixed resolution is a $\gamma$ distance of less than $0.05$ from the main plateau 91\% of the time for the period 1991--1998, 93\% of the time for 1999--2003, and 88\% of the time for 2005--2008. The community configurations of the main plateau and the fixed resolution differ in the community assignments of fewer than five nodes in 78\% of time-steps for the period 1991--1998, in 83\% of time-steps for 1999--2003, and in 88\% of time-steps for 2005--2008. For the majority of time-steps, the community configuration at the fixed resolution is hence identical or very similar to the configuration corresponding to the main plateau. This supports our proposed method of investigating the community dynamics at a fixed $\gamma$ for each period.

\subsection{Testing community significance} \label{sub::sigtest}

The scaled energy (see Eq.~\ref{Q_POTTS}) measures the strength of communities compared with some null model, so large scaled energies indicate more significant communities. To ensure that the identified communities are meaningful, we perform a permutation test \cite{GOOD_BOOK_2005} and compare the scaled energies of the observed community partitions with the scaled energies for community partitions derived from shuffled data. For the period 1991--2003, we generate shuffled data for each of the USD exchange rates by randomly reordering the returns of the corresponding time series. We create shuffled data for each of the non-USD exchange rates using the shuffled USD time series and the triangle relations described in Section \ref{sec::data}. We then calculate new correlation matrices for these shuffled time series, form new adjacency matrices and find the communities and scaled energies for each of the new networks. Similarly for the period 2005--2008, we shuffle the returns for each of the exchange rates in the set \{AUD/USD, EUR/NOK, EUR/SEK, EUR/USD, GBP/USD, NZD/USD, USD/CAD, USD/CHF, USD/JPY, USD/XAU\} and calculate the return time series for each of the rates not in this set by applying the triangle relations to these shuffled time series. This procedure conserves the return distribution for each of the original USD exchange rates for the period 1991--2003 and for each of the rates in the above set for 2005--2008. The shuffling, however, destroys the temporal correlations. Any structure in the shuffled data therefore emerges as a result of the triangle relationships. The  shuffled data therefore provides some insights into the effects of the triangle relations on the properties of the actual data.

\begin{figure}
\includegraphics[width=0.85\linewidth]{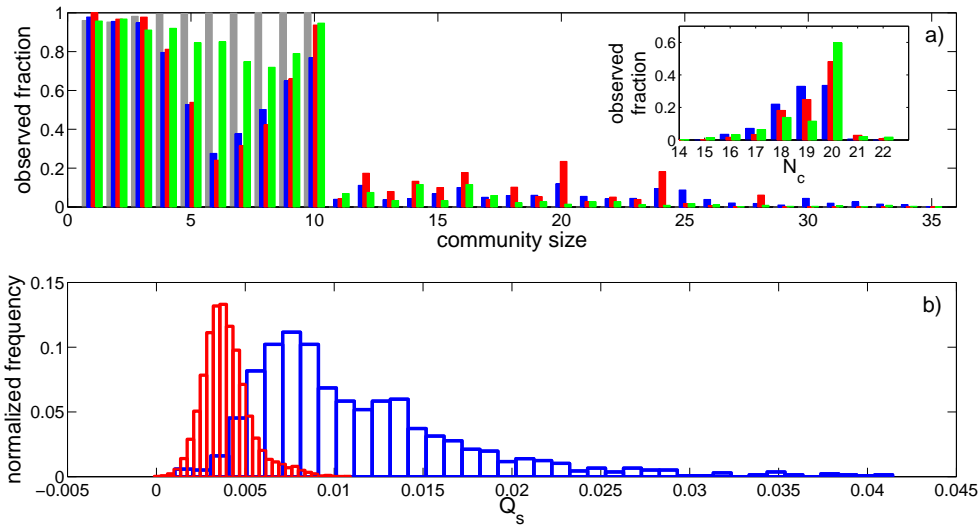}
\caption{\label{commStats}(Color online)
(a) The fraction of time-steps at which a community of a given size is observed for 1991--1998 (blue), 1999--2003 (red), 2005--2008 (green), and shuffled data (gray). The shuffled data distribution combines the results for the period 1991--2003 and for 2005--2008. The distributions were almost identical for the two periods. The inset shows the fraction of time-steps at which $N_c$ communities are observed for 1991--1998 (blue), 1999-2003 (red), and 2005--2008 (green). (b) Comparison of the distribution of the scaled energy for 1991--2003 for market data (blue) and 100 realizations of shuffled data (red).}
\end{figure}

By inspection, Fig.~\ref{commStats}(b) shows that the communities identified for the actual data are significantly stronger than those generated using shuffled data. The sample mean scaled energy for the actual data is $0.011$ (with a standard deviation of $0.0061$) and for shuffled data the sample mean is $0.0039$ (with a standard deviation of $0.0013$). The communities observed for the actual data are therefore significantly stronger than the communities for randomized data in which the structure results from the triangle effect. This provides strong evidence that the communities represent meaningful structures within the FX market, so these communities can provide insights into the correlation structure of the market. We now consider properties of these communities in detail.
%We compare the distributions in Fig.~\ref{commStats}(b) by performing a two-sample Kolmogorov-Smirnov test \cite{STATS_BOOK} for the null %hypothesis that the two samples are drawn from the same distribution. The test rejects the null hypothesis at the 1\% significance level, with 
%a $p$-value of $p=0$.
%an asymptotic $p$-value computed numerically to be $0$, indicating that the null hypothesis is rejected even under the most critical conditions. %Therefore, we can conclude that the scaled energies for the real data are not drawn from the same distribution as the scaled energies for %shuffled data and that the observed communities are 
%in general 
%significantly stronger than those generated by randomized data.

% do we state explicitly that asymptotic p-value computed using kstest2 in MATLAB?

\subsection{Community properties} \label{sub::properties}

The inset of Fig.~\ref{commStats}(a) shows that the number of communities into which the network is partitioned exhibits only small fluctuations from 1991--2008. Nevertheless, as shown in Fig.~\ref{platStats}(c), there is a considerable variation in the extent of community reorganization between consecutive time-steps.  No nodes change community between some steps, whereas more than twenty nodes change communities between others. Figure~\ref{commStats}(a) shows that the community size distribution is bimodal for all three periods, with a tail extending to large community sizes. There is therefore a large variation in the sizes of the communities observed at each time-step for all three periods. However, the minimum between the two peaks is not as deep for the period 2005--2008 and has shifted from a community size of $6$ nodes to one of $8$.

The peak in the size distribution for communities with 10 members occurs as a result of the the number of currencies in the network. For each of the eleven currencies XXX$\in$\{AUD, CAD, CHF, GBP, DEM, JPY, NOK, NZD, SEK, USD, XAU\}, there are ten exchange rates XXX/YYY with XXX as the base currency (and ten equivalent inverse rates YYY/XXX). We derive most of the exchange rates in a set of rates with the same base currency by applying the triangle relation (see Section~\ref{sec::data}) to pairs of exchange rate time series; one of the rates is common across across all of the exchange rates in the base currency set and the other rate is different for each rate in the set. For example, for the period 1991--2003, we derive the CAD/DEM exchange rate from the USD/CAD and USD/DEM rates while we derive the CAD/GBP rate from the USD/CAD but with the USD/GBP. Exchange rates with the same base currency are, therefore, often correlated and consequently they have a tendency to form communities with ten members. However, it is not possible for all currencies to form a ten-member base currency community at each time-step. 
%Consider a time-step at which there is a community of ten gold exchange rates of the form XAU/YYY. There will also be an inverse community of ten exchange rates of the form XXX/XAU in the opposite half of the network. The existence of this inverse community means that it is not possible for there to be any other communities including ten exchange rates with the same base currency. For example, there cannot be a community of ten exchange rates of the form USD/YYY because the USD/XAU rate is in the XXX/XAU community. At most nine USD/YYY rates can belong to the same community. 
If there is no additional structure beyond these base-currency correlations that emerge as a result of the triangle relation, then one would expect to observe communities with $1,2,\cdots,10$ members at each time-step (and equivalent communities of inverse rates). Figure \ref{commStats}(a) shows that this size distribution is indeed observed for shuffled data. However, Fig.~\ref{commStats}(a) also shows that the community size distribution for market data is significantly different, so the community detection techniques are uncovering additional FX market correlations. This again demonstrates that the triangle effect is not dominating the results.

The frequently-observed communities shown in Table \ref{commEg} demonstrate the variation in community size. Some of the most common communities are single exchange rates, such as USD/CAD, which are formed of two closely-related currencies. Table \ref{commEg} also highlights that communities often consist of exchange rates with the same base currency. In \cite{MM_PRE_2008}, McDonald et al.\ used the relative clustering strengths of groups of exchange rates with the same base currency to provide insights into the effects of important news and events on individual currencies. The relative size of different base-currency communities can provide similar information. For example, if we observe a community of ten CHF/YYY exchange rates and a community of three DEM/YYY, the larger size of the CHF/YYY community suggests that the CHF is more important than the DEM in the market at this time.

It is also worth noting that the most frequently observed community of ten exchange rates with the same base currency is the gold (XAU) community.  We include gold in our study because there are many similarities between it and a currency. However, gold also tends to be more volatile than most currencies, so it is unsurprising that the gold rates often form their own community.  Consequently, the absence of a large gold community at a time-step is often a good indication that another currency is particularly influential.

Importantly, the identified communities do not always contain exchange rates with the same base currency, providing insights into changes in the inherent values of different currencies. For example, consider a community containing several exchange rates with CHF as the base currency and several rates with DEM as the base currency. The fact that the exchange rates are in the same community suggests that they are correlated. The structure of this community also provides information about the inherent values of the CHF and DEM. Exchange rates of the form XXX/YYY quote the value of one currency in terms of another currency, so if the price of XXX/YYY increases it is not clear whether this is because XXX has become more valuable or because YYY has become less valuable. However, if one observes that the price of XXX increases against a range of different YYY over the same period, then one expects that the value of XXX has increased. Therefore, returning to our example, if one observes a community of several CHF/YYY and DEM/YYY exchange rates for many different YYY then this suggests that these rates are positively correlated. Because the values of CHF and DEM have increased against a range of other currencies, we expect that the inherent values of both CHF and DEM are increasing.

\begin{table}[ht]
\caption{Examples of frequently-observed communities for the pre-euro period 1991--1998 and for the two post-euro periods (1999--2003 and 2005--2008). The quantity Fr denotes the fraction of time-steps at which each community is observed.}
\begin{center}\footnotesize
\begin{tabular}{cp{8.3cm}c}
\hline
\hphantom{0}Period\hphantom{00} &Community &Fr\\
\hline
\multirow{9}{*}{1991--1998} &USD/CAD &0.62\\
&DEM/CHF &0.45\\
&NZD/\{CAD,USD\} &0.33\\
&AUD/\{CAD,NZD,USD\} &0.32\\
&XAU/\{AUD,CAD,CHF,DEM,GBP,JPY,NOK,NZD,SEK,USD\} &0.28\\
&SEK/\{AUD,CAD,CHF,DEM,GBP,JPY,NOK,NZD,USD,XAU\} &0.17\\
&DEM/NOK &0.16\\
&AUD/\{CAD,NZD,USD,XAU\} &0.14\\
&GBP/\{CHF,DEM,NOK\} &0.12\\
\hline \hline
\multirow{9}{*}{1999--2003} &EUR/CHF &0.88\\
&USD/CAD &0.67\\
&XAU/\{AUD,CAD,CHF,EUR,GBP,JPY,NOK,NZD,SEK,USD\} &0.64\\
&NOK/\{CHF,EUR\} &0.59\\
&SEK/\{CHF,EUR,NOK\} &0.51\\
&GBP/\{CAD,USD\} &0.24\\
&NZD/\{AUD,CAD,CHF,EUR,GBP,JPY,NOK,SEK,USD\} &0.21\\
&JPY/\{CAD,GBP,USD\} &0.17\\
&AUD/\{CAD,CHF,EUR,GBP,JPY,NOK,SEK,USD\} &0.14\\
\hline \hline
\multirow{9}{*}{2005--2008} &XAU/\{AUD,CAD,CHF,EUR,GBP,JPY,NOK,NZD,SEK,USD\} &0.91\\
&EUR/CHF &0.65\\
&AUD/NZD &0.39\\
&CAD/\{AUD,CHF,EUR,GBP,JPY,NOK,NZD,SEK,USD\} &0.39\\
&GBP/\{CHF,EUR\} &0.35\\
&SEK/\{CHF,EUR\} &0.33\\
&NZD/\{AUD,CAD,CHF,EUR,GBP,JPY,NOK,SEK,USD\} &0.26\\
&NOK/\{CHF,EUR,SEK\} &0.21\\
&GBP/\{CHF,EUR,NOK,SEK\} &0.20\\
\hline \hline
\end{tabular}
\label{commEg}
\end{center}
\end{table}

%%%%%%%%%%%%%%%%%%%%%%%%%%%%%%%%%%%%%%%%%%%%

\section{Minimum spanning trees} \label{sec::mst}

Perhaps the best-known approach for studying networks of financial assets is to consider the minimum spanning tree (MST) of the network. MSTs have been used regularly in studies of equity markets to identify clusters of stocks that belong to the same market sector \cite{JPO_PRE_2003,JPO_EPJ_2004,MAN_EPJ_1999,BOU_BOOK_2003}. In this section, we briefly consider the limitations of this approach for community detection and describe the additional information that the Potts method can provide.

MSTs are constructed using the agglomerative hierarchical clustering technique known as single-linkage clustering \cite{DUDA_BOOK_2001,MAP_SURVEY_2009}. Agglomerative methods start with $n$ singleton clusters and create a hierarchy by sequentially linking clusters based on their similarity. At the first step, the two nodes separated by the smallest distance are joined in a cluster. At each subsequent step, the distance between the new cluster and each of the old clusters is recomputed and the two clusters again joined. This can be repeated until all clusters are connected. The similarity of clusters $c$ and $c'$ is usually expressed as a distance $\mathcal{D}(c,c')$, which is determined by considering the distance $d_{ij}$ between each node $i\in{c}$ and each node $j\in{c'}$. In single-linkage clustering, the distance between clusters is given by
\begin{equation}
	\mathcal{D}(c,c') = \min_{\begin{subarray}{1} i\mspace{2mu}\in{c}\\ j\in{c'} \end{subarray}}\thickspace{d_{ij}}\,.
\end{equation}
Single-linkage clustering thus respresents an extreme because it joins clusters based on the minimum distance between nodes in each cluster. An alternative is average-linkage clustering, for which
\begin{equation}
	\mathcal{D}(c,c') = \frac{1}{|c||c'|}\sum_{i\in{c}}\sum_{j\in{c'}}d_{ij}\,.
\end{equation}
%where $\vert\cdot\vert$ indicates the number of nodes in community $c$.

For financial networks, the standard measure used for $d_{ij}$ is the nonlinear transformation of the correlation coefficient $\rho(i,j)$ given by \cite{MAN_EPJ_1999,MAN_ECON_2000}
\begin{equation}
	d_{ij}=\sqrt{2(1-\rho(i,j))}. \label{ud}
\end{equation}
The distance takes values $d_{ij}\in[0,2]$ with small values between similar nodes. The MST directly gives rise to the subdominant ultrametric hierarchical organization of the nodes. This ultrametric space was originally chosen as a suitable space in which to link nodes in a study of a network of equities because it was observed a posteriori to give rise to a financially-meaningful taxonomy of assets \cite{MAN_EPJ_1999,JPO_PRE_2003}.

In constructing MSTs, the merging of clusters $c$ and $c'$ corresponds to adding an edge between the closest nodes in $c$ and $c'$. The edges must always link clusters, so that the network never has any closed loops. If the agglomeration is continued until there is a path from every node to every other node, one obtains a spanning tree. Because the clusters are joined using the minimum distance between pairs of nodes, MSTs necessarily possess the minimum total edge length of any possible spanning tree. A minimum spanning tree is, therefore, a simply connected, acyclic graph that connects the $n$ nodes in a network with $n-1$ links, which is appealing because its $n-1$ links make it much simpler to analyze than the full network with $\frac{1}{2}n(n-1)$ links. An alternative representation of the output of a linkage clustering algorithm, which shows the full hierarchical structure, is a dendrogram (or hierarchical tree) \cite{DUDA_BOOK_2001,MAP_SURVEY_2009}. At the first level of the dendrogram, there are $n$ singleton clusters. As one climbs the vertical distance scale of the dendrogram, clusters are combined consecutively until all nodes are contained in a single community at the top of the dendrogram.

\begin{figure}
\includegraphics[width=0.85\linewidth]{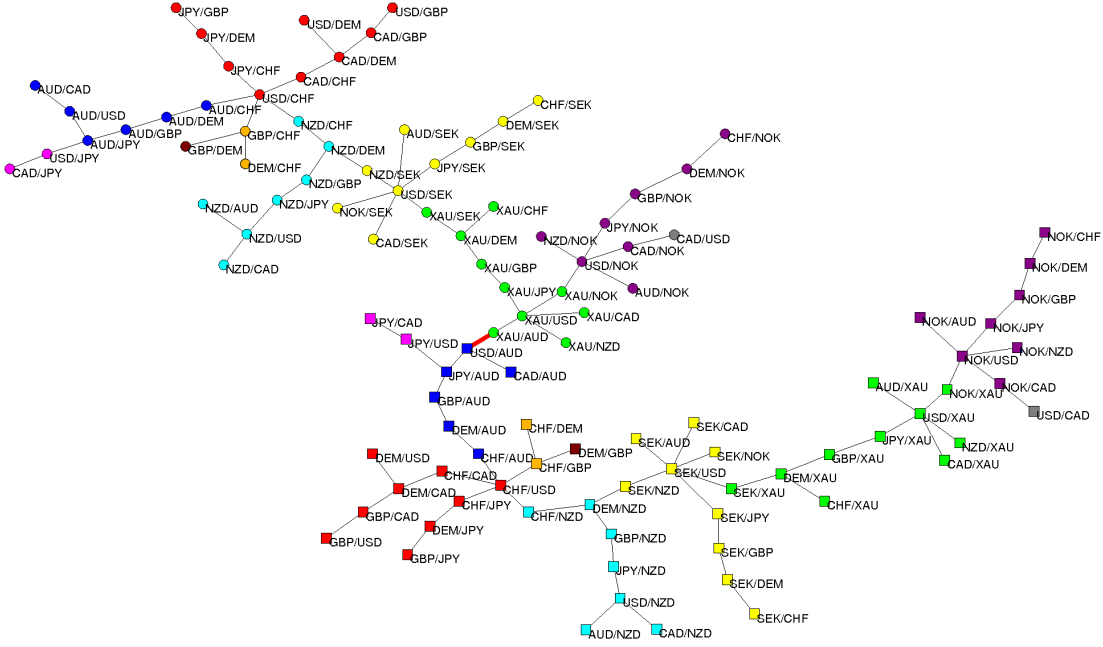}
\caption{\label{mst}(Color online) The minimum spanning tree for the network formed from a time window of returns beginning on 09/18/1991. The tree is split into two identical halves (indicated by $\bigcirc$ and $\square$), which are connected via the edge (shown in red) between the XAU/USD and USD/AUD exchange rates. For each community of exchange rates, there is an equivalent community of inverse rates in the other half of the tree. We color each node according to its community membership determined using the Potts method with $\gamma=1.41$, and we show each community of exchange rates in the same color as the corresponding community of inverse rates.}
\end{figure}

In earlier studies of equity markets, clusters of closely-related assets were identified by their proximity on the branches of the MST \cite{JPO_PRE_2003,JPO_EPJ_2004,MAN_EPJ_1999} and by finding the disconnected groups of assets that remained when all tree links weaker than some threshold were removed \cite{BOU_BOOK_2003}. Similar analyses have been performed that find clusters of assets by considering the whole network and removing edges below some threshold or alternatively by starting with a network with no links and iteratively adding links above an increasing threshold \cite{JPO_EPJ_2004,GARAS_EPJB_2008}. In Fig.~\ref{mst}, we show an example of an MST of exchange rates. We color the nodes in this tree according to their community membership as determined using the Potts method. The MST is partitioned into two halves with communities of exchange rates in one half and equivalent communities of inverse exchange rates in the other. In this example, nodes belonging to the same community are always linked in the MST, but this is not always the case. 

The main problem with single-linkage clustering (and, as a consequence, with MSTs) is that clusters can be joined as a result of single pairs of elements being close to each other even though many of the elements in the two clusters have large separations. The MST then contains weak links that might be misinterpreted as being more financially meaningful than they actually are \cite{JPO_EPJ_2004}. It is also difficult to determine where the community boundaries lie on the MST. For example, a branch of an MST might include nodes belonging to a single community or the nodes might belong to several communities. As an example of this phenomenon, and of the additional clustering information provided by the Potts method, consider the branch at the far right of the tree shown in Fig.~\ref{mst}. By simply considering the MST, one might have inferred the existence of a cluster that includes all of the NOK/YYY rates and USD/CAD. However, the Potts method highlights the fact that USD/CAD forms a singleton community and that NOK/XAU belongs to a community with the XXX/XAU rates. This observation might provide information as to the relative importance of NOK and XAU in the market over this period.

\begin{figure}[htp]
\begin{center}
\subfigure[]{\label{dendsingle}\includegraphics[width=0.85\linewidth]{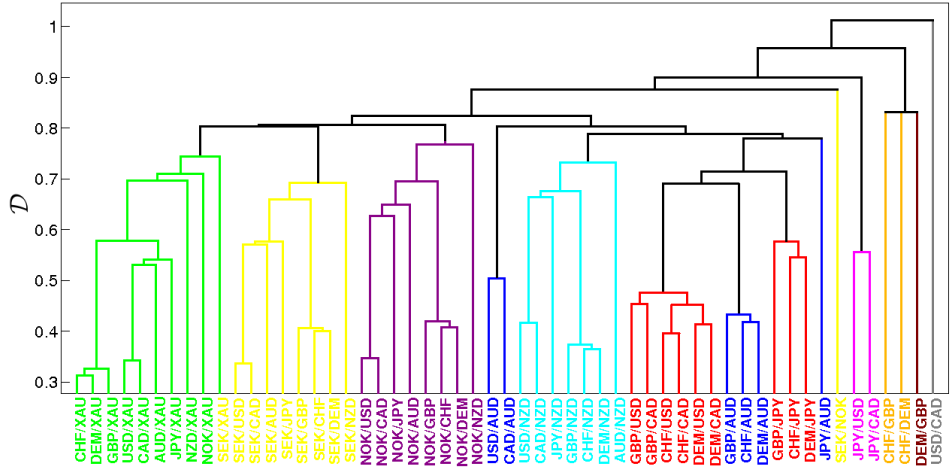}}\\
\subfigure[]{\label{dendave}\includegraphics[width=0.85\linewidth]{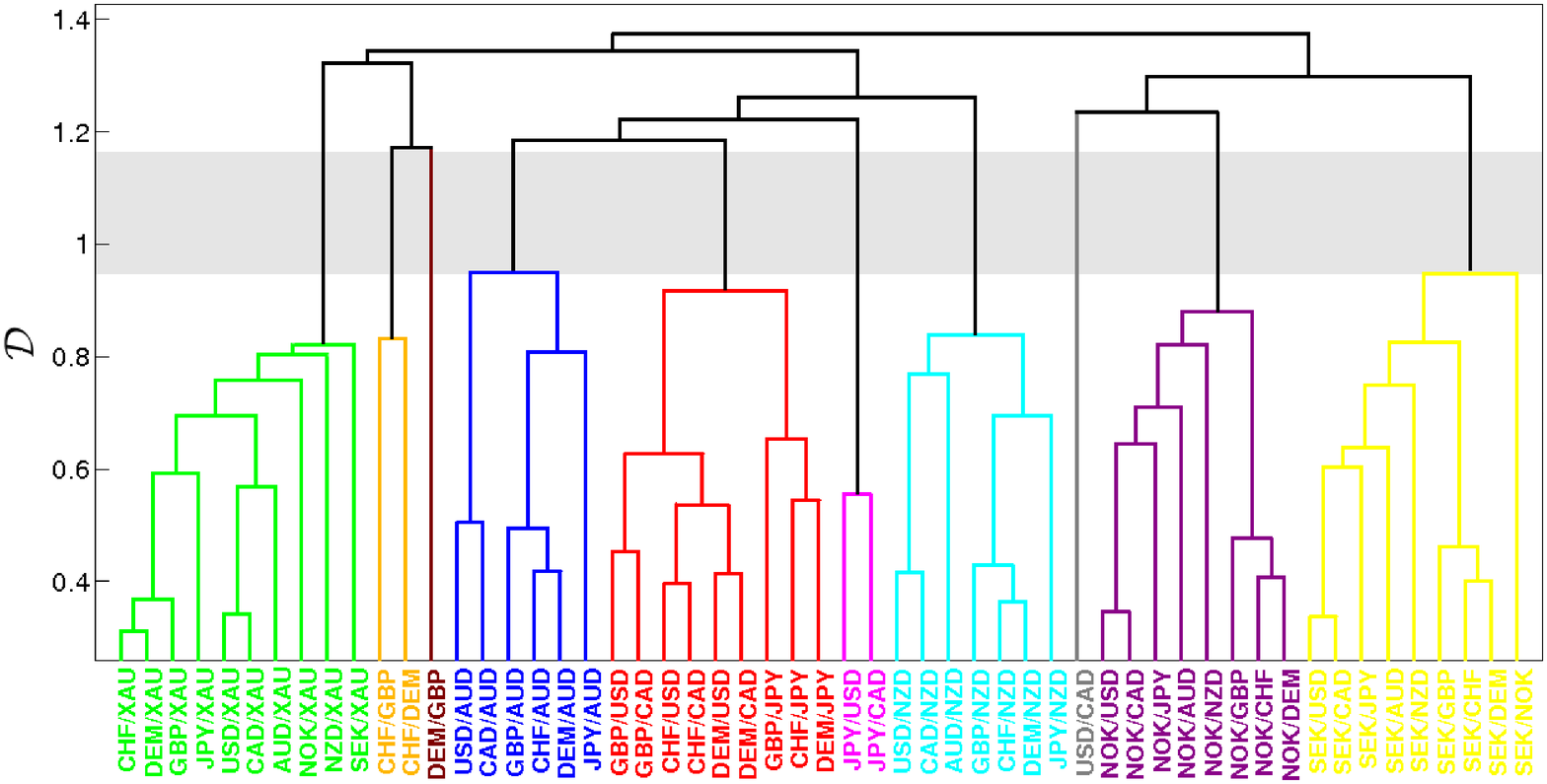}}
\end{center}
\caption{(Color online) Dendrograms showing the hierarchical clustering of exchange rates for one half of the network for a time window of returns beginning on 09/18/1991. We color each exchange rate according to its community membership determined using the Potts method with $\gamma=1.41$. We generated the dendrograms using (a) single linkage clustering and (b) average linkage clustering. The gray region in panel (b) highlights the range over which the communities correspond to the communities of the main plateau identified using the Potts method.}
\label{dend}
\end{figure}

In Fig.~\ref{dend}(a), we show the dendrogram generated using the same single-linkage clustering algorithm used to produce the MST in Fig.~\ref{mst}. If the distances between different dendrogram levels are reasonably uniform, then no clustering appears more ``natural'' than any other \cite{DUDA_BOOK_2001}. However, large distances between levels (i.e., the same clusters persist over a large range of distances) might indicate the most appropriate level at which to view the clusters. This is analogous to investigating communities that are robust over a range of resolutions. The clusterings observed at some levels of Fig.~\ref{dend}(a) correspond quite closely with the communities identified using the Potts method, but there is no level at which they correspond exactly. The levels are also reasonably evenly distributed along the distance axis. In the dendrogram in Fig.~\ref{dend}(b), which was generated using average-linkage clustering, there is a range of distances over which the clustering does not change. The clustering observed over this interval is identical to the community configuration corresponding to the main plateau found using the Potts method. Therefore, in this case, average-linkage clustering and the Potts method identify the same robust communities.

%%%%%%%%%%%%%%%%%%%%%%%%%%%%%%%%%%%%%%%%%%%%

\section{Exchange rate centralities and community persistence} \label{sec::centroles}

Thus far, we have considered the properties of entire communities. We now investigate the roles of nodes within communities.

\subsection{Centrality measures} \label{sub::centralities}

We describe the relationship between a node and its community using various centrality measures. In the social networks literature, such measures are used to measure the roles of nodes within the network and to identify which nodes are the most important or most prominent \cite{WASSERMAN_BOOK_2008}. Because there are multiple notions of importance, a multitude of different centrality measures have been proposed \cite{VAL_CONN_2008}. In the present context, we use centrality measures to identify exchange rates that occupy important positions within the FX market.

The \textit{betweenness centrality} of nodes is defined using the number of geodesic paths between pairs of vertices in a network \cite{FREE_SOCI_1977,NEW_SIAM_2003}. We calculate node betweenness by taking the distance between nodes $i$ and $j$ as
\begin{equation}
	d_{ij}=
		\begin{cases}
			0 &\text{if $i=j$ or $A_{ij}=1$}\,,\\
			1/A_{ij} &\text{otherwise}\,.
		\end{cases}
\end{equation}
The betweenness centrality $b_i$ of node $i$ is then given by
\begin{equation}
	b_i=\sum_s\sum_t\frac{g^i_{st}}{G_{st}},\qquad\textrm{for }{s,t}\neq{i}\textrm{ and }s\neq{t},
\end{equation}
where $G_{st}$ is the total number of shortest paths from node $s$ to node $t$ and $g_{st}^i$ is the number of shortest paths from $s$ to $t$ passing through $i$. Betweenness centrality is widely used in social network analysis to quantify the extent to which people lie on paths that connect others. Nodes with high betweenness can be considered to be important for facilitating communication between others in the network, so betweenness is used to help measure the importance of nodes for the spread of information around the network \cite{VAL_CONN_2008}.

We also consider the {community centrality} of each node \cite{NEW_PRE_2006}. We employ the scaled energy matrix $\mathbf{J}$, with components $J_{ij}=A_{ij}-\gamma{P_{ij}}$, where we again set $P_{ij}=k_ik_j/2m=(n-2)/2n$. Following the notation in \cite{NEW_PRE_2006}, the energy matrix can be expressed as $\mathbf{J}=\mathbf{UDU}^T$, where $\mathbf{U}=(\mathbf{{u_1}\vert{u_2}\vert\cdots})$ is the matrix of eigenvectors of $\mathbf{J}$ and $\mathbf{D}$ is the diagonal matrix of eigenvalues $\beta_i$. If $\mathbf{D}$ has $q$ positive eigenvalues, then one can define a set of node vectors $\{\mathbf{x}_i\}$ of dimension $q$ by
\begin{equation}
	[\mathbf{x}_i]_j=\sqrt{\beta_i}U_{ij},\qquad{j\in\{1,2,...,q\}}\,,
\end{equation}
where $[\mathbf{x}_i]_j$ indicates the $j$th ($j\in{1,\cdots{q}}$) element of the node vector of node $i$. The magnitude $\vert\mathbf{x}_i\vert$ is the \textit{community centrality}. Nodes with high community centrality play an important role in their local neighborhood, irrespective of the community boundaries.

One can also define a community vector
\begin{equation}
	\mathbf{w}_k=\sum_{i\in{c}^k}\mathbf{x}_i
\end{equation}
for each community $k$ with members $c^k$. Nodes with high community centrality are strongly attached to their community if their node vector is also aligned with their community vector. Continuing to use the definitions in \cite{NEW_PRE_2006}, a \textit{projected community centrality} $y_i$ is defined by
\begin{equation}
	y_i=\mathbf{x}_i\cdot\hat{\mathbf{w}}_k=\vert\mathbf{x}_i\vert\cos{\theta_{ik}}
\end{equation}
and we refer to the quantity $\cos{\theta_{ik}}$ as the \emph{community alignment}. The community alignment is near $1$ when a node is at the center of its community and near $0$ when it is on the periphery. Nodes with high community alignment are located near the center of their community and have a high projected community centrality, so they are strongly attached to their community and can be considered to be highly influential within it. The number of positive eigenvalues of $\mathbf{J}$ can vary between time-steps, so we normalize $\vert\mathbf{x}_i\vert$ and $y_i$ by their maximum value at each time-step.

\subsection{Community tracking} \label{sub::tracking}

A node's identity is known at all times and its community is known at any one time. We can thus track community evolution from the perspective of individual nodes. We investigate the persistence through time of nodes' communities by defining a community autocorrelation. For a node $i$ with community $c_i(t)$ at time $t$, the autocorrelation $a_i^t(\tau)$ of its community after $\tau$ time-steps is defined by
\begin{equation}
	a_i^t(\tau)=\frac{\mid{c}_i(t)\cap{c}_i(t+\tau)\mid}{\mid{c}_i(t)\cup{c}_i(t+\tau)\mid}\,. \label{nodecentric}
\end{equation}
This is a node-centric version of a quantity considered in \cite{VICSEK_Nat_2007} and importantly does not require one to determine which community at each time-step represents the descendant of a community at the previous time-step. In \cite{VICSEK_Nat_2007}, Palla et al.\ detect communities using a method known as clique percolation and then track communities by finding the community at $t+1$ that has the maximum edge overlap with a community at $t$. Several other approaches have been proposed for identifying community descendants using different measures to quantify the node-overlap rather than the edge-overlap between communities at different time-steps \cite{BERGERWOLF_2006,ASUR_2007}. In \cite{BERGERWOLF_2006}, Berger-Wolf and Saia assume that two communities are similar if their node overlap (measured using the Jaccard distance \cite{TRAUD_2009}) exceeds some threshold. Similarly in \cite{ASUR_2007}, Asur et al.\ identify split and merged communities based on a function of the union and intersection of the nodes in the descendant and parent communities exceeding a threshold.

In \cite{FALKOWSKI_2006}, Falkowski et al.\ identify communities using modularity maximization and again match communities at different time-steps based on the node overlap exceeding some threshold. However, instead of simply comparing communities at consecutive time-steps, they match communities between all times that are within $\tau$ time-steps of each other. They then construct a graph where each community observed over the full evolution of the network represents a node and connect all nodes for which the community overlap exceeds the selected threshold (where only communities within $\tau$ time-steps of each other are compared). Finally they find groups of similar communities by running a community detection algorithm on this new network.

Toyoda and Kitsuregawa \cite{TOYODA_2003} track communities using a method that avoids the use of an arbitrary threshold by identifying the descendant of community $c(t)$ as the community $c(t+1)$ that shares the largest number of nodes with $c(t)$. If multiple communities share the same number of nodes, they select the community with the largest total number of nodes as the descendant.

Methods that identify descendant communities based on maximum node or edge overlap can, however, lead to equivocal mappings following splits and mergers. For example, consider a community $c^f(t)$ that splits into two communities $c^g(t+1)$ and $c^h(t+1)$ at the following time-step. If the overlap between $c^f(t)$ and $c^g(t+1)$ is identical to the overlap between $c^f(t)$ and $c^h(t+1)$ then one will need to make an arbitrary choice as to which community represents the descendant of $c^f(t)$. In {\cite{WANG_ARX_2008}} Wang et al.\ propose a method that deals with the issue of splits and mergers by tracking communities using ``core'' nodes. They find core nodes by aggregating the difference between the degree of each node and each of its nearest neighbors. A core node is then any node that has an aggregate degree that is greater than or equal to zero and communities are tracked from the perspective of these nodes. However, this approach then requires that for a community to be considered to persist it must contain the same core node at multiple time- steps, which seems overly restrictive. In order to avoid all of the above ambiguities, instead of tracking whole communities we identify communities from the perspective of individual nodes.

\begin{figure}
\includegraphics[width=0.85\linewidth]{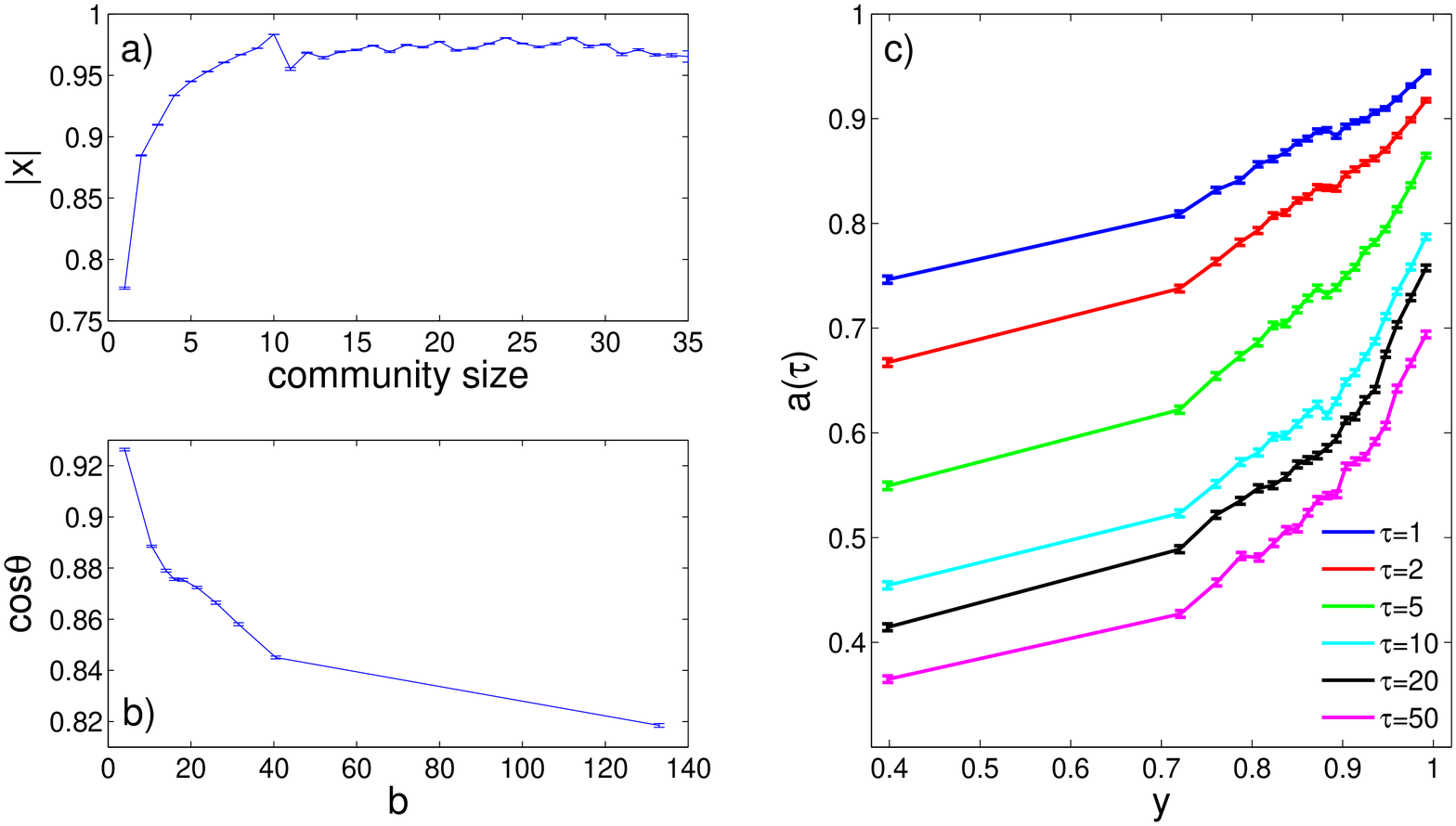}
\caption{\label{station}(Color online) (a) Mean community centrality versus the size of the community to which the node belongs. (b) Mean community alignment versus the betweenness centrality of nodes. (c) Mean community autocorrelation versus the projected community centrality. (All error bars indicate the standard error \cite{STATS_BOOK}.)
}
\end{figure}

\subsection{Exchange rate roles} \label{sub::erroles}

In Fig.~\ref{station}(a), we show the mean normalized community centrality of exchange rates as a function of community size (averaging over all nodes belonging to the same-size community). The community centrality increases with community size up to sizes of about $10$ members. For larger communities, $\vert\mathbf{x}_i\vert$ remains approximately constant. Nodes with high $\vert\mathbf{x}_i\vert$ therefore tend to belong to large communities, so exchange rates with high community centrality tend to be closely linked with many other rates. Table \ref{table:centralities} shows the ten exchange rates that tend to have the highest betweenness centrality, community centrality, and projected community centrality. For all three periods, CHF/NZD, CHF/XAU, and SEK/XAU have one of the ten highest community centralities, so they are closely tied to many other rates. For 1991--2003, exchange rates formed from one of the major European currencies---DEM (and then EUR, after its introduction) or CHF---and one of the commodity currencies\footnote{A country is said to have a ``commodity currency" if its export income depends heavily on a commodity. For example, AUD, NZD and CAD are all considered to be commodity currencies.} also tend to have high community centrality.  For 2005--2008, however, XAU rates encompass nearly all of the exchange rates with the highest $\vert\mathbf{x}_i\vert$.

Figure \ref{station}(b) shows the mean betweenness centrality versus the community alignment. We calculate the mean community position by splitting the range of $b$ into $10$ bins containing equal numbers of data points and then averaging over all community positions falling within these bins. (The observed relationships are robust for reasonable variations in the number of bins.) Nodes with high betweenness centrality tend to have small values for their community position, implying that nodes that are important for information transfer are usually located on the edges of communities. Table \ref{table:centralities} shows that for all three periods, NOK/SEK, AUD/NZD, and AUD/CAD all tend to have high betweenness centrality on average.  They are therefore located on the edges of communities and are important for information transfer.  Interestingly, for the post-euro period (1999--2008), several USD exchange rates are also important for information transfer, but no USD rates regularly have high betweenness for the pre-euro period. In contrast, XAU exchange rates are important for information transfer for the pre-euro period but not after the euro was introduced.

In Fig.~\ref{station}(c), we show the mean community autocorrelation versus the projected community centrality. We calculate the mean autocorrelation by splitting the range of $y$ into $15$ bins containing equal numbers of data points and then averaging over all autocorrelations falling within these bins. (Again, the observed relationships are robust for reasonable variations in the number of bins.) As one would expect, the community autocorrelation for the projected community centrality of a given node is smaller for larger $\tau$. More interesting is that for all values of $\tau$, the mean community autocorrelation increases with $y$. This suggests that nodes that are strongly connected to their community are likely to persistently share that community membership with the same subset of nodes. In contrast, exchange rates with a low $y$ experience regular changes in the set of rates with which they are clustered.

Table \ref{table:centralities} shows the exchange rates with the highest projected community centrality, which in turn reveals the most persistent communities. For 1991--2003, approximately half of the ten exchange rates with the highest projected community centrality also appear in the list of the ten rates with the highest community centrality.  For 2005--2008, however, the lists of exchange rates with the highest community centrality and projected community centrality are dominated by the same set of XAU exchange rates (though the rankings differ). For 1991--2003, the exchange rates with the highest projected community centrality again includes rates formed of DEM (and EUR) or CHF and one of the commodity currencies.  However, there are also a number of USD exchange rates with high projected community centrality that don't have high community centrality. This suggests that these USD rates do not have strong links with a large number of other exchange rates, but that they strongly influence the rates within their community.

\begin{table}[ht]
\caption{The ten exchange rates with the highest betweenness centrality, community centrality, and projected community centrality for each of the three periods. We rank the exchange rates for each centrality according to their average rank over all time-steps. For each exchange rate XXX/YYY the equivalent inverse rate YYY/XXX had the same betweenness centrality, community centrality, and projected community centrality.}
\begin{center}\footnotesize
\begin{tabular}{@{}cccccccccccc@{}}
\hline
\hphantom{0}\multirow{2}*{Rank}\hphantom{0} &\multicolumn{3}{c}{1991--1998} & &\multicolumn{3}{c}{1999--2003} & &\multicolumn{3}{c}{2005--2008}\\
\cline{2-4}\cline{6-8}\cline{10-12}
&$b$ &$\vert{\mathbf{x}}\vert$ &$y$ & &$b$ &$\vert{\mathbf{x}}\vert$ &$y$ & &$b$ &$\vert{\mathbf{x}}\vert$ &$y$\\
\hline
\hphantom{0}1 &NOK/SEK &CHF/AUD &USD/DEM & &AUD/NZD &SEK/XAU &USD/XAU & &USD/CAD &JPY/XAU &EUR/XAU\\
\hphantom{0}2 &AUD/XAU &CHF/NZD &USD/CHF & &NZD/CAD &CHF/CAD &EUR/USD & &AUD/NZD &USD/XAU &GBP/XAU\\
\hphantom{0}3 &AUD/NZD &CHF/XAU &USD/XAU & &AUD/CAD &EUR/XAU &EUR/XAU & &AUD/CAD &NZD/XAU &CHF/XAU\\
\hphantom{0}4 &AUD/CAD &CHF/CAD &CHF/CAD & &JPY/CAD &NOK/XAU &GBP/XAU & &NOK/SEK &CAD/XAU &EUR/CAD\\
\hphantom{0}5 &CHF/SEK &DEM/AUD &CHF/AUD & &NOK/SEK &CHF/NZD &EUR/CAD & &USD/GBP &GBP/XAU &SEK/XAU\\
\hphantom{0}6 &NZD/XAU &SEK/AUD &CHF/NZD & &USD/AUD &CHF/XAU &USD/CHF & &NZD/CAD &SEK/XAU &USD/XAU\\
\hphantom{0}7 &CAD/XAU &DEM/XAU &DEM/CAD & &USD/NZD &EUR/CAD &CHF/XAU & &USD/JPY &CHF/XAU &EUR/NZD\\
\hphantom{0}8 &DEM/SEK &SEK/XAU &DEM/AUD & &USD/JPY &EUR/NZD &NOK/XAU & &USD/AUD &NOK/XAU &JPY/XAU\\
\hphantom{0}9 &NZD/CAD &NOK/AUD &USD/AUD & &GBP/JPY &SEK/NZD &EUR/NZD & &CHF/NOK &CHF/NZD &AUD/XAU\\
\hphantom{0}10 &DEM/NOK &DEM/NZD &DEM/NZD & &CHF/SEK &NOK/NZD &CHF/NZD & &GBP/AUD &AUD/XAU &NOK/XAU\\
\hline
\end{tabular}
\label{table:centralities}
\end{center}
\end{table}

%%%%%%%%%%%%%%%%%%%%%%%%%%%%%%%%%%%%%%%%%%%%

\section{Major community changes} \label{sec::majorchange}

\begin{figure}[ht]
\begin{center}
\centerline{\includegraphics[width=0.85\linewidth]{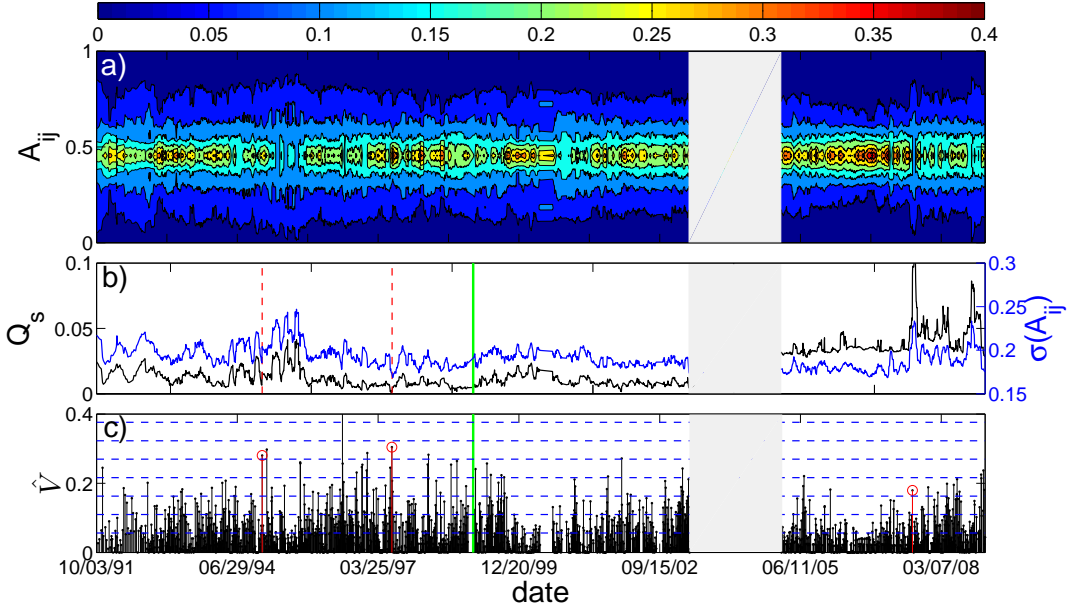}}
\caption{\label{evolution}(Color online)
(a) Normalized distribution of the link weights at each time-step. (b) Scaled energy $Q_s$ and standard deviation of the link weights. (c) Normalized variation of information $\hat{V}$ between the community configurations at consecutive time-steps. The horizontal lines show (from bottom to top) the mean of $\hat{V}$ and $1$, $2$, $3$, $4$, $5$, and $6$ standard deviations above the $\hat{V}$ mean. The green vertical line in panels (b) and (c) separates the pre- and post-euro periods. The red vertical lines show the time-steps when 12/22/94, 02/07/97, and 08/15/07 enter the rolling time window. These dates correspond, respectively, to the devaluation of the Thai baht during the Asian currency crisis, the flotation of the Mexican peso following its sudden devaluation during the tequila crisis, and significant unwinding of the carry trade during the 2007--2008 credit crisis. The gray blocks mark 2004 (for which we had no data).}
\end{center}
\end{figure}

We now investigate the insights that short-term community dynamics can provide into changes in the FX market. Figure \ref{evolution}(a) shows a contour plot of the normalized distribution of the link weights at each time-step. The mean link strength remains constant through time because of the inclusion in the network of each exchange rate and its inverse but [as one can see in Figs.~\ref{evolution}(a,b)] there is a large variation in the standard deviation of the link strengths. The scaled energy and standard deviation of link weights are closely related. This is expected because the standard deviation increases as a result of the strengthening of strong ties and the weakening of weak ones.

In Fig.~\ref{evolution}(c), we also show $\hat{V}$ between the community configurations at consecutive time-steps. Large spikes in $\hat{V}$ indicate significant changes in the community configuration over a single time-step and potentially also indicate important market changes. The correlation coefficient between $\hat{V}$ and the absolute change in $Q_s$ between consecutive time-steps is 0.39 over the period 1991--2003 and 0.47 over the period 2005--2008. The correlation between $\hat{V}$ and the absolute change in $\sigma(A_{ij})$ is 0.28 over the period 1991--2003 and 0.27 from 2005--2008. Changes in $Q_s$ are thus a better indicator than changes in $\sigma(A_{ij})$ that there has been a change in the community configuration of the network.

\begin{figure*}[ht]
\begin{center}
\centerline{\includegraphics[width=0.85\linewidth]{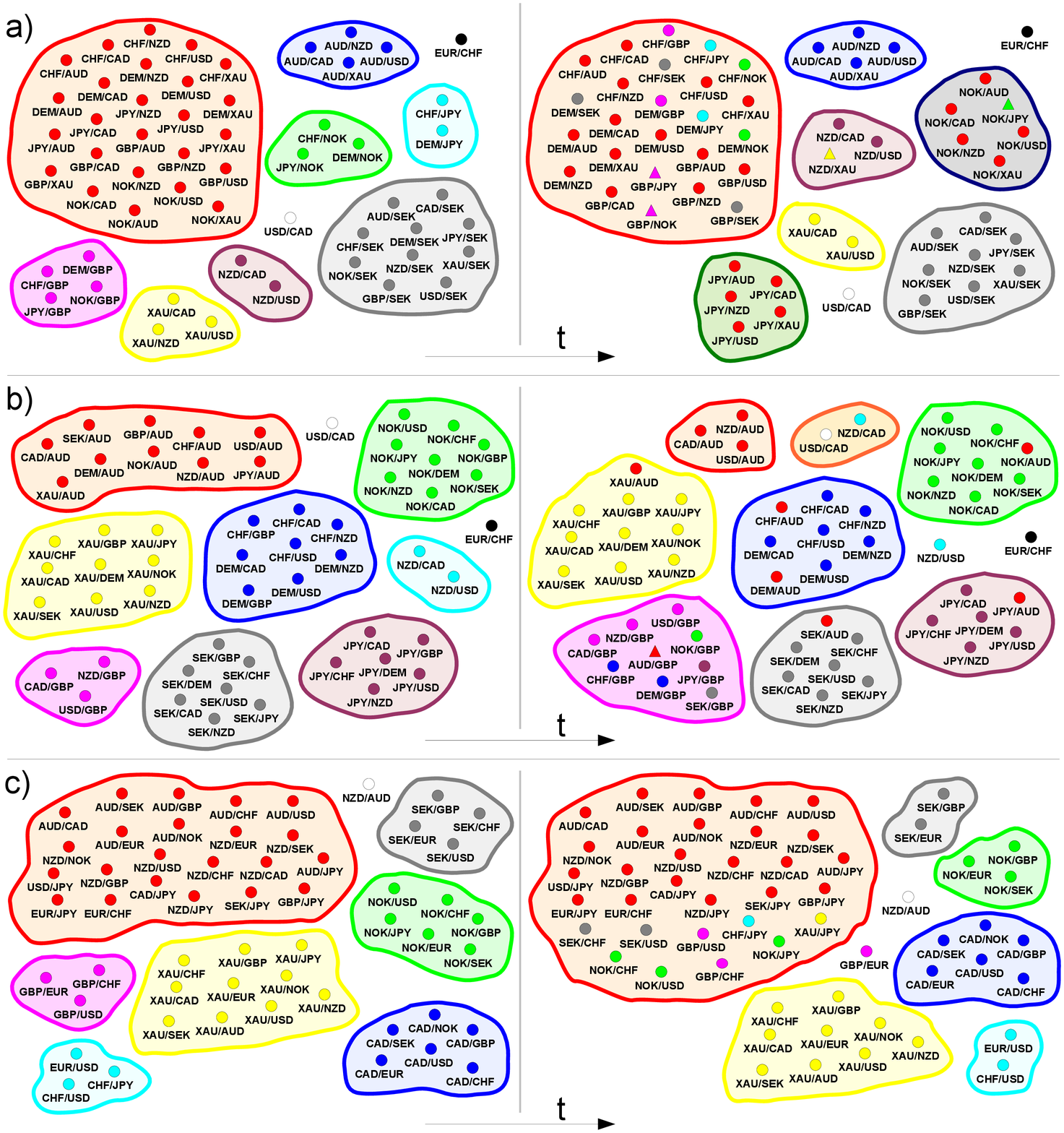}}
\caption{\label{comm_schematic}(Color online)
Schematic representation of the change in the community structure in one half of the FX market network for several events.  (a) The Mexican tequila crisis: the depicted reorganization followed 12/22/94, when the Mexican peso was allowed to float after a sudden devaluation. (b) The Asian currency crisis: the depicted reorganization followed 07/02/1997, when Thailand devalued the baht. (c) Carry trade unwinding: the depicted reorganization followed 08/15/07, when there was significant unwinding of the carry trade during the 2007--2008 credit and liquidity crisis. The node colors after the community reorganization correspond to their community before the change. If the parent community of a community after the reorganization is obvious, we draw it using the same color as its parent. The nodes drawn as triangles resided in the opposite half of the network before the community reorganization.}
\end{center}
\end{figure*}

In Fig.~\ref{comm_schematic}, we show three example community reorganizations---two in which $\hat{V}$ is more than four standard deviations larger than its mean and a third in which it is over two standard deviations above the mean.

\subsection{Mexican peso crisis}
Figure \ref{comm_schematic}(a) shows the reorganization on 12/22/1994 when the Mexican peso was floated following its sudden devaluation.\footnote{For a floating exchange rate, the value of the currency is allowed to fluctuate according to the FX market. Prior to its floatation the peso had been pegged to the US dollar.} This change is accompanied by an increase in the scaled energy. Although we do not include the Mexican peso in the set of investigated exchange rates, it appears that its devaluation was a sufficiently serious event to cause major changes in the community relationships of the studied rates. Before 12/22/1994, the largest community consisted of a group of exchange rates of the form \{AUD, CAD, NZD, USD, XAU\}/\{CHF, DEM, GBP, JPY, NOK\}. After the floatation, the largest community consisted of a set of exchange rates formed from the major European currencies (CHF, DEM, and GBP). It is also noteworthy that there is only a small gold (XAU) community during this period which, as noted previously, often indicates that another currency is particularly important in the market.

\subsection{Asian currency crisis}
Figure \ref{comm_schematic}(b) shows the community changes following 07/02/1997 when the Thai baht was devalued during the Asian currency crisis. As with the peso, although we did not include the baht in the set of studied rates, its devaluation appears to have had a significant effect on the whole market. There is a large stable gold cluster during the whole period. Before 07/02/1997, there is also a large AUD cluster, but after the devaluation, this cluster breaks up and the previously-small GBP cluster increases in size. This suggests that the GBP is playing a more prominent market role after the devaluation. Although the reasons for the changes in the sizes of the AUD and GBP communities are not obvious, both adjustments suggest a sharp and significant change in the correlation structure of the market.

\subsection{Credit crisis}
\label{subsec:carry}
The final example in Fig. \ref{comm_schematic}(c) shows the community changes following 08/15/07 when there was a significant community reorganization and reveals one of the major effects on the FX network of the recent credit and liquidity crisis.
 %when there was significant unwinding of the carry trade during the recent credit and liquidity crisis.  
This example also demonstrates community changes that occurred as a result of a trading change that affected the studied rates directly. 
%The period 2005--2008 includes the recent credit and liquidity crisis. 

The most important effect of the credit crisis on the FX market during the period 2005--2008 was its impact on the carry trade. The carry trade consists of selling low interest rate ``funding currencies'' such as the JPY and CHF and investing in high interest rate ``investment currencies'' such as the AUD and NZD. It yields a profit if the interest rate differential between the funding and investment currencies is not offset by a commensurate depreciation of the investment currency \cite{BRUNNERMEIER_2008}. The carry trade is one of the most commonly used FX trading strategies and requires a strong appetite for risk, so the trade tends to ``unwind" during periods in which there is a decrease in available credit. A trader unwinds a carry trade position by selling his/her holdings in investment currencies and buying funding currencies.

\begin{figure}[ht]
\begin{center}
\centerline{\includegraphics[width=0.85\linewidth]{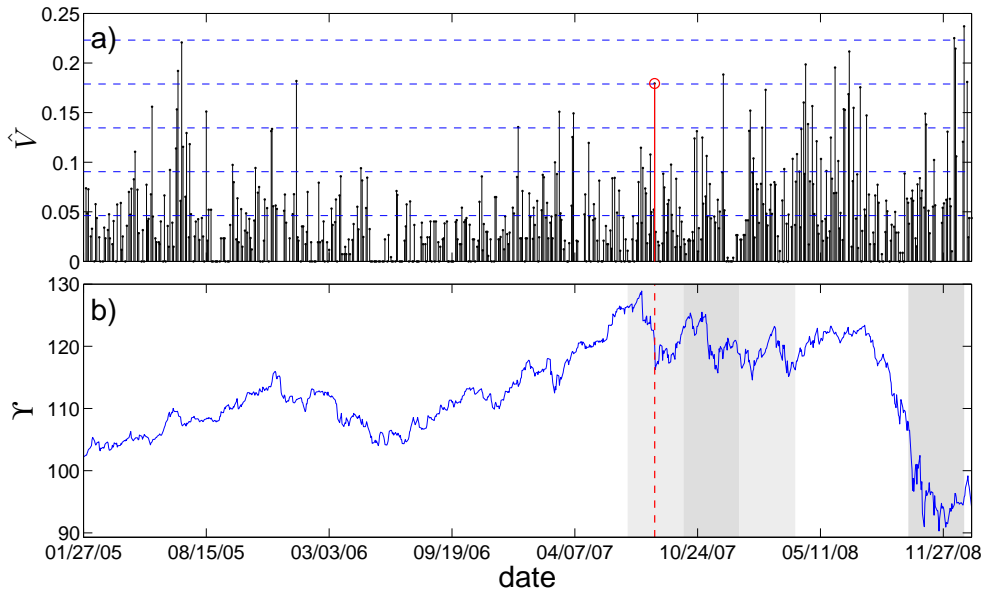}}
\caption{\label{carry}(Color online)
(a) Normalized variation of information between the community configuration at consecutive time steps for 2005--2008. The horizontal lines show (from bottom to top) the mean of $\hat{V}$ and $1$, $2$, $3$, and $4$ standard deviations above the mean. The red vertical line in (a) shows the 08/15/07 when there was a marked increase in unwinding of the carry trade. (b) Carry trade index $\Upsilon$. The vertical line again shows 08/15/07 and the shaded blocks (from left to right) Q3 2007, Q4 2007, Q1 2008, and Q4 2008.}
\end{center}
\end{figure}

One approach to quantifying carry trade activity is to consider the returns that can be achieved using a carry trade strategy. In Fig. \ref{carry}(b) we show the cumulative return index $\Upsilon$ from trading a common carry trade strategy. We consider a strategy in which one buys equal weights of the three major currencies with the highest interest rates and sells equal weights of the three currencies with the lowest interest rates. This is a dynamic trading strategy because the relative interest rates of currencies change over time. For example, consider the situation in which the interest rate of currency $A$ (which initially has the third highest interest rate) decreases below the rate of currency $B$ (which initially has the fourth highest interest rate). In order to maintain the strategy of only holding the three currencies with the highest interest rates at any time, one would re-balance the carry portfolio by selling the holding of currency $A$ and buying currency $B$. The frequency at which such re-balances occur will depend on the frequency at which the relative interest rates change. The returns from a carry strategy like this are widely seen by market participants to provide a good gauge of carry trade activity. Large negative returns result in large decreases in $\Upsilon$ which are therefore likely to indicate significant unwinding of the carry trade.

In Fig.~\ref{carry}(a) we focus on the period 2005--2008 from Fig.~\ref{evolution}(c). Again, large spikes indicate significant changes in the community configuration over a single time-step. Figure \ref{carry}(a) shows that a significant community reorganization occurred on 08/15/07 and in Fig.~\ref{comm_schematic}(c) we show the observed communities before and after this date. This community change is a result of massive unwinding of the carry trade. Figure \ref{carry}(b) shows that leading up to 08/15/07 there was some unwinding of the carry trade so the initial configuration includes a community containing exchange rates of the form AUD/YYY, NZD/YYY, and XXX/JPY (which all involve one of the key carry-trade currencies). In Fig. \ref{carry}(b) it is also clear that following this date there is a sharp increase in carry trade unwinding. The second community partition in Fig. \ref{comm_schematic}(c) highlights this increase as the carry community increases in size by incorporating other XXX/JPY rates as well as some XXX/CHF and XXX/USD rates. The presence of a large number of exchange rates involving one of the key carry-trade currencies in a single community clearly demonstrates the significance of the trade over this period. Importantly, some of the exchange rates included in the carry community are also somewhat surprising and provide insights into the range of currencies used in the carry trade over this period.

The analysis above demonstrates that one can identify major changes in the correlation structure of the FX market by finding large values of $\hat{V}$ between time-steps. Having identified significant changes, one can gain a better understanding of the nature of these changes and potentially also gain insights into trading changes taking place in the market by investigating the adjustments in specific communities.  We have discussed three examples in which the observed changes are obviously attributable to a major FX market event. However, there are also a number of time-steps when significant community reorganizations occur for which the cause is much less obvious, and the analysis of dynamic communities might help shed light on related market changes.

\section{Visualizing changes in exchange rate roles} \label{sec::rolevis}

In this final section we investigate changes in the relationships between specific exchange rates and their communities. We begin by defining within-community $z$-scores, which directly compare the relative importances of different nodes to their community \cite{GUIM_NAT_2005}. We describe the roles of individual nodes at each time-step using the within-community projected community centrality $z$-score $z^y$ and within-community betweenness centrality $z$-score $z^b$. If a node $i$ belongs to community $c_i$ and has projected community centrality $y_i$, then
\begin{equation}
	z^{y}_i=\frac{y_i-\bar{y}_{c_i}}{\sigma^{y}_{c_i}}\,,
\end{equation}
where $\bar{y}_{c_i}$ is the mean of $y$ over all nodes in $c_i$ and $\sigma^{y}_{c_i}$ is the standard deviation of $y$ in $c_i$. The quantity $z^{y}_i$ measures how strongly connected node $i$ is to its community compared with other nodes in the same community. Similarly, if the same node has betweenness centrality $b_i$, then
\begin{equation}
	z^b_i=\frac{b_i-\bar{b}_{c_i}}{\sigma^b_{c_i}}\,,
\end{equation}
where $\bar{b}_{c_i}$ is the average of $b$ over all nodes in $c_i$ and $\sigma^{b}_{c_i}$ is the standard deviation of $b$ in $c_i$. The quantity $z^b_i$ indicates the importance of node $i$ to the spread of information compared with other nodes in its community.\footnote{Note that in order for a within-community $z$-score to be well defined, a node must belong to a community containing two or more nodes.} The positions of nodes in the $(z^b,z^y)$ plane thereby illuminate the roles of the associated exchange rates in the FX market and provide information that cannot be gained by simply considering individual exchange rate time series.

We remark that our methods are robust with respect to the choice of measures used to construct the parameter plane:  we obtain similar results using other notions, such as dynamical importance \cite{REST_PRL_2006} instead of the betweenness centrality and the within-community strength $z$-score \cite{GUIM_NAT_2005} instead of the projected community centrality.

\subsection{Average roles} \label{sub::averole}

\begin{figure}[htp]
\begin{center}
\subfigure{\label{allRoles_pre}\includegraphics[width=0.85\linewidth]{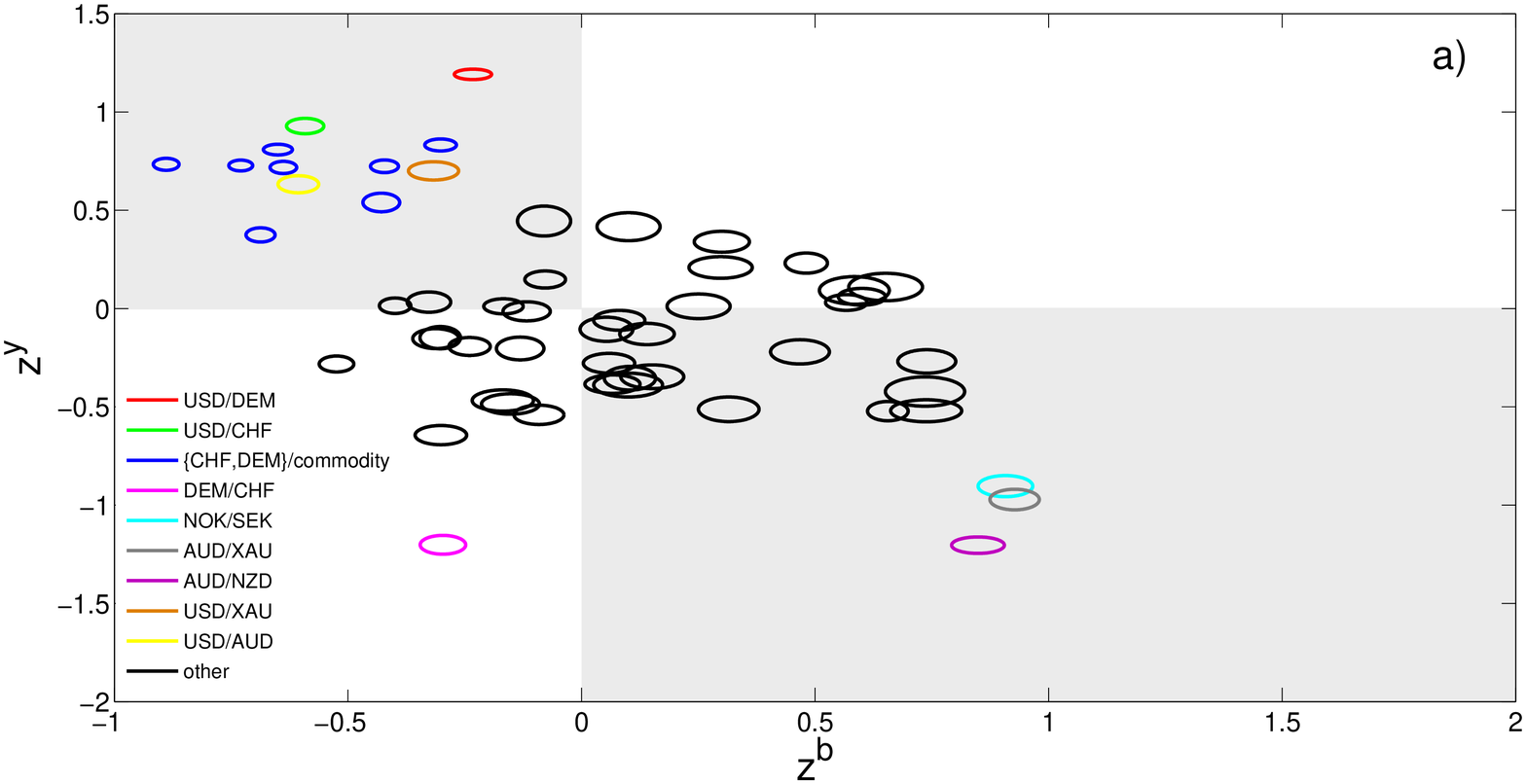}}\\
\subfigure{\label{allRoles_post}\includegraphics[width=0.85\linewidth]{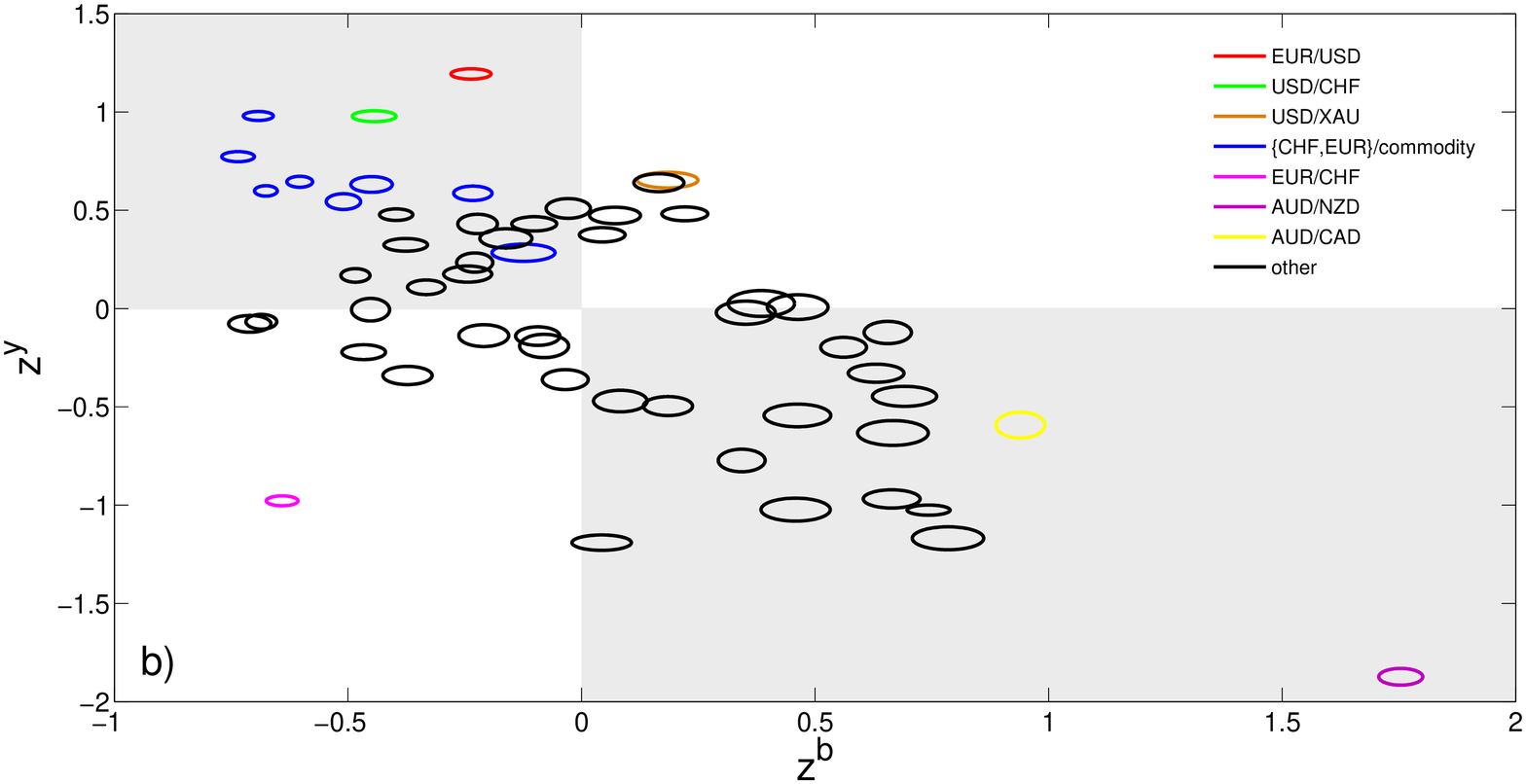}}\\
\subfigure{\label{allRoles_crunch}\includegraphics[width=0.85\linewidth]{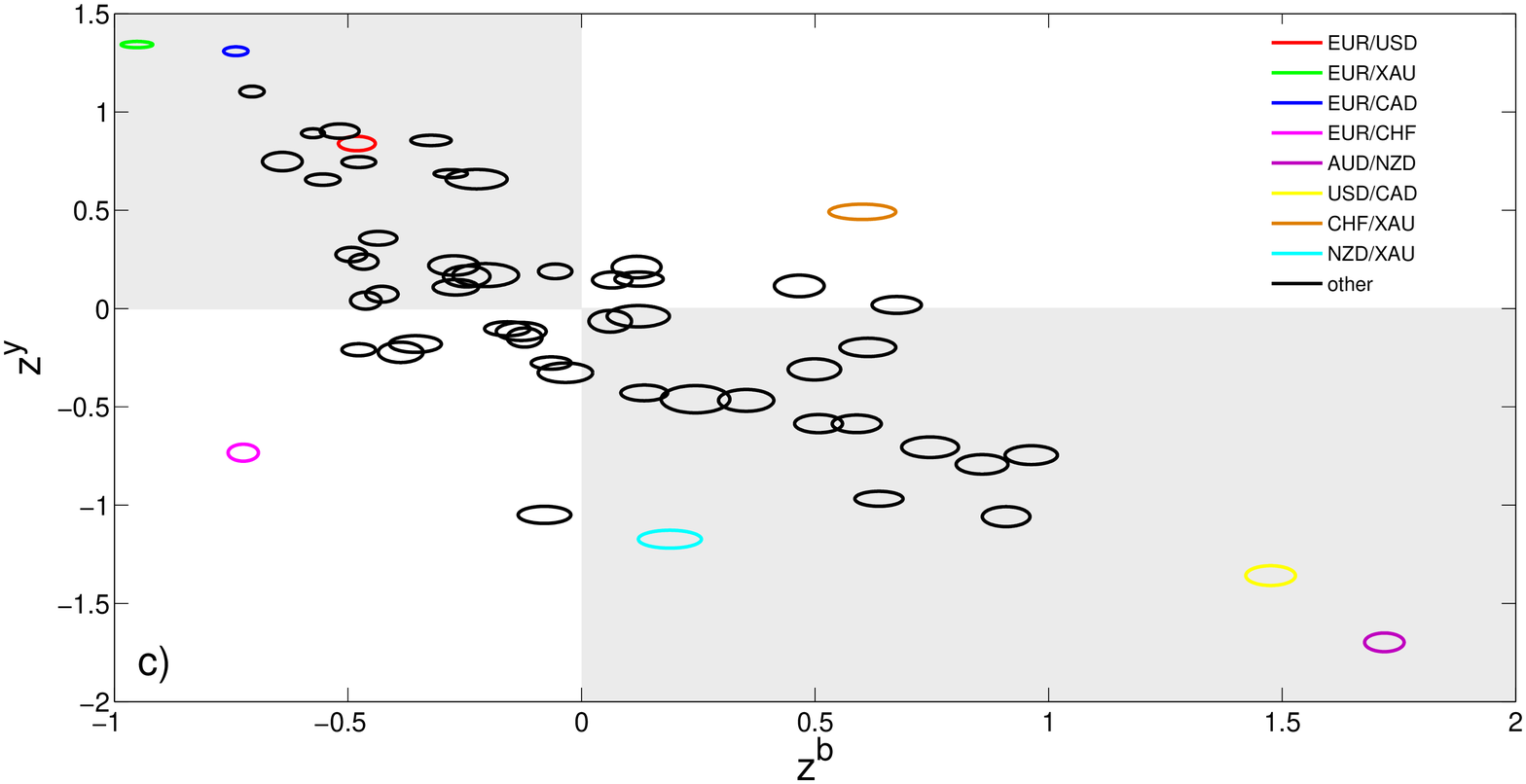}}
\end{center}
\caption{(Color online) Node positions in the $(z^b,z^y)$ plane averaged over all time-steps for the periods (a) 1991--1998, (b) 1999--2003, and (c) 2005--2008. The radii of each elliptical marker equal the standard deviations in the parameters for the corresponding node scaled by a factor of $1/15$ for visual clarity.
}
\label{allRoles}
\end{figure}

In Fig.~\ref{allRoles}, we show the mean position of each exchange rate over the three periods and highlight some rates that play particularly prominent roles. For example, the USD/DEM (and then EUR/USD after the introduction of the euro) regularly had the strongest connection to its community from 1991-2003, but EUR/XAU was more strongly connected to its community from 2005--2008. The importance of USD/DEM and EUR/USD is unsurprising given that these rates had the highest daily trading volume \cite{BIS_2001}. This provides a reality check that our methods uncover useful information about the roles of minor exchange rates. Other exchange rates, such as NOK/SEK and AUD/NZD, were less influential within their communities but were very important for the transfer of information around the network.

The $(z^b,z^y)$ plots also highlight exchange rates that play similar roles in the FX market. For example, exchange rates formed from one of the major European currencies---DEM or CHF---and one of the commodity currencies---AUD, CAD, and NZD (or the commodity XAU)---are located close together in the upper left quadrant of the $(z^b,z^y)$ plane for 1991--2003. This prominent similarity is not present for 2005--2008.

\subsection{Annual roles} \label{sub::annrole}

\begin{figure}[ht]
\begin{center}
\centerline{\includegraphics[width=0.85\linewidth]{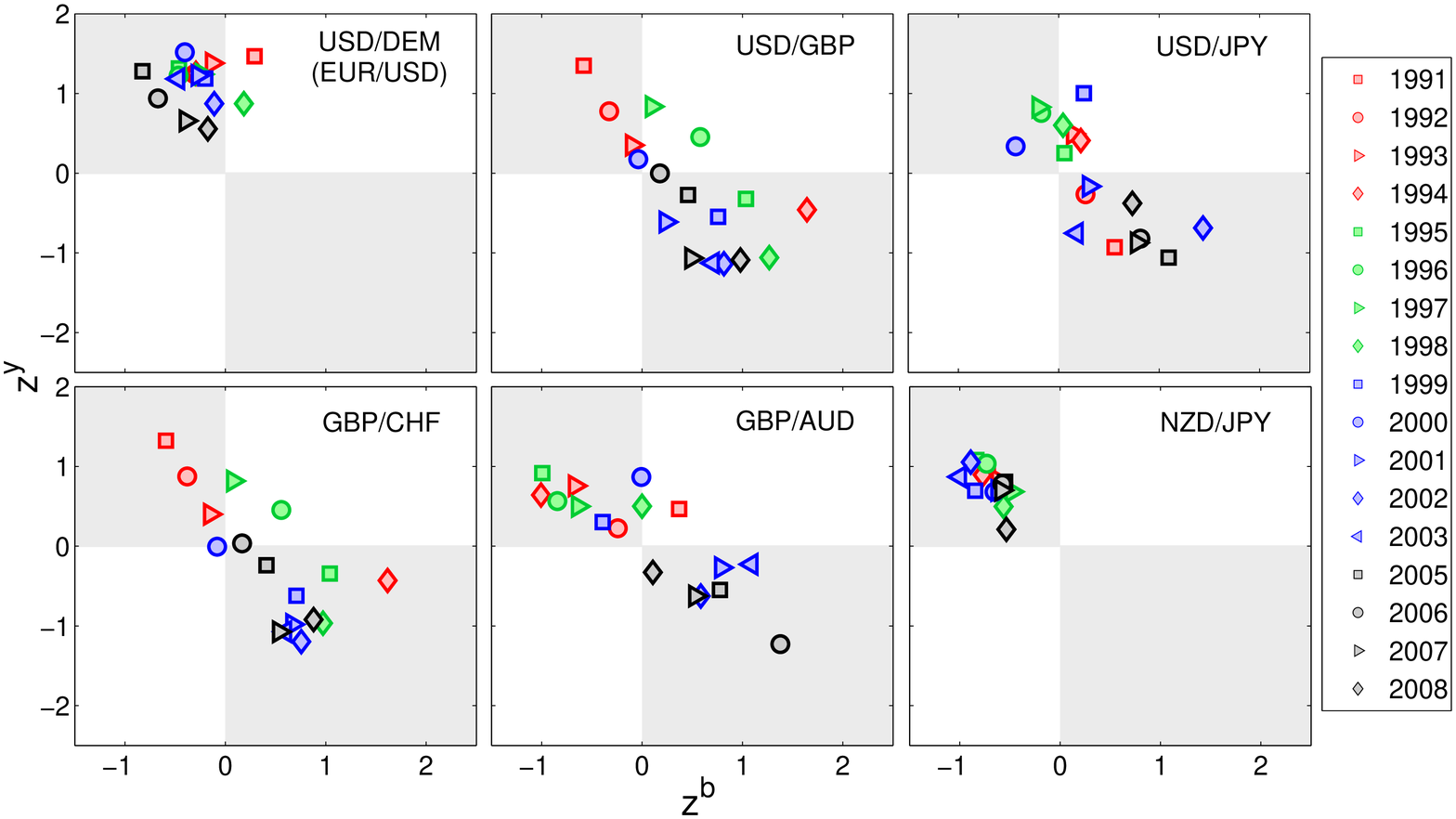}}
\caption{\label{roles}(Color online) Annual node role evolutions in the $(z^b,z^y)$ plane for the full period 1991--2008.}
\end{center}
\end{figure}

We can also gain insights into the time dynamics of exchange rate roles by examining changes in the positions of the rates in the $(z^b,z^y)$ plane over different time periods. Changes in a node's position in the $(z^b,z^y)$ plane reflect changes in the membership of a node's community as well as changes in $b$ and $y$. In Fig.~\ref{roles}, we show six example annual role evolutions.  We determine the annual roles by averaging $z^y$ and $z^b$ over all time-steps in each year.  We see, for example, that the NZD/JPY exchange rate maintained a consistently influential role within its community over the full period and similarly the EUR/USD rate also maintained the same influential role played by the USD/DEM rate before the introduction of the euro.

Other rates changed role over the studied period. The GBP/USD and GBP/CHF exchange rates evolved in a similar manner, as they changed from being strongly influential within their communities before 1994 to being less influential within their communities but more important for information transfer after 1994.  The role of both GBP/AUD and USD/JPY varied significantly from 1991--2008: From 2001 onwards the GBP/AUD became less influential within its community but more important for information transfer. Interestingly, the USD/JPY had its highest within-community influence in the late 1990s during a period of Japanese economic turmoil. One can construct similar plots to study the change in the role of other exchange rates. These role plots provide a useful tool for visualizing the changes in the exchange rate correlations.

\subsection{Quarterly roles} \label{sub::quartrole}

\begin{figure}[ht]
\begin{center}
\centerline{\includegraphics[width=0.85\linewidth]{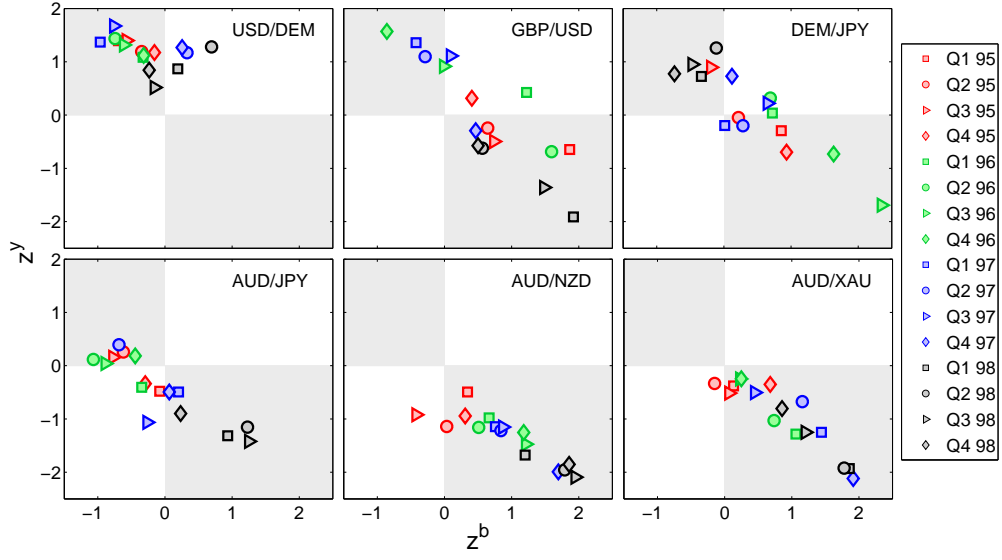}}
\caption{\label{quarterlyRoles9598}(Color online) Quarterly node role evolutions in the $(z^b,z^y)$ plane for the period 1995--1998.}
\end{center}
\end{figure}

We also investigate higher-frequency changes in exchange rate market role over shorter time intervals. In Fig.~\ref{quarterlyRoles9598} we show quarterly roles changes over the period 1995--1998 for six exchange rates including USD/DEM and GBP/USD for which we also show the annual changes in Fig.~\ref{roles}. Both exchange rates show similar role variations over both time-scales, with the USD/DEM always playing a relatively influential role within its community and the GBP/USD role varying significantly. We also show other examples for which we did not show annual changes. The role of DEM/JPY varied considerably from 1995--1998: In particular, it was an important information carrier for the last two quarter in 1996, but was influential within its community throughout 1998. In contrast the AUD/JPY move from being unimportant for information transfer to being an information carrier during 1998. In further contrast, the AUD/NZD and AUD/XAU were both always information carriers to different degree with the AUD/NZD being particularly important for information transfer during 1998.

\begin{figure}
\includegraphics[width=0.85\linewidth]{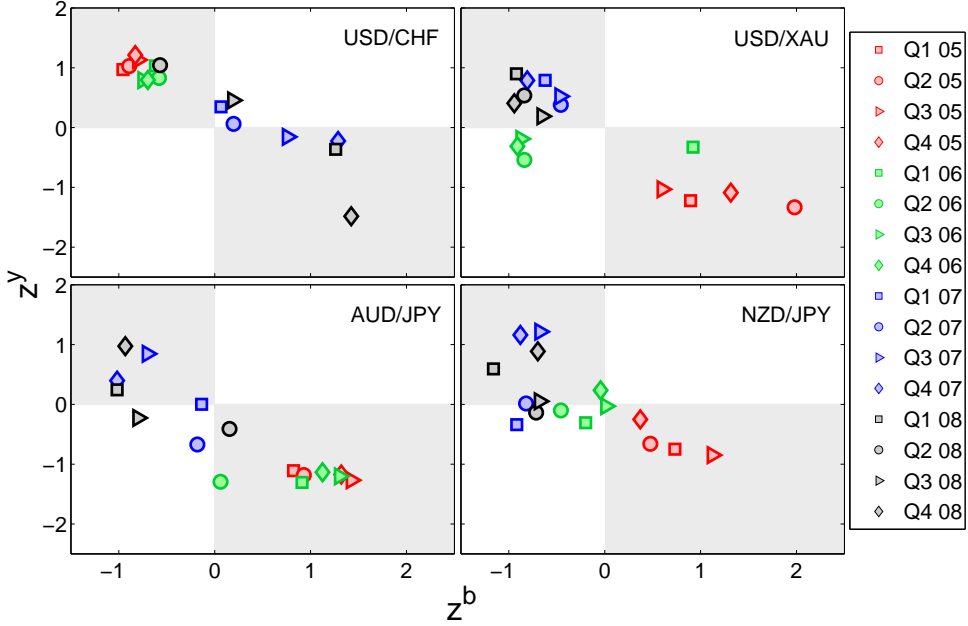}
\caption{(Color online) Quarterly node role evolutions in the $(z^b,z^y)$ plane for the period 2005--2008.}\label{quarterlyroles}
\end{figure}

Finally, we consider some examples of quarterly role evolutions for the period 2005--2008 which we discussed in our recent short paper \cite{FENN_2009}. Figure~\ref{quarterlyroles}, shows quarterly role changes for four exchange rates from 2005--2008. The USD/XAU rate provides an interesting example due to the persistence of its community over this period. From 2005--2008, the USD/XAU node shifted from being an important information carrier within the XAU community to being more central to this community. This period of higher influence coincides closely with the period of financial turmoil during 2007--2008. The CHF is widely regarded as a ``safe haven'' currency \cite{RANALDO_2008}, so one might expect USD/CHF to behave in a similar manner to USD/XAU. However, the CHF is also a key carry trade currency. Because CHF is used both as a safe haven and as a carry trade currency, the USD/CHF node does not move in the same direction as USD/XAU in the $(z^b,z^y)$ plane. Instead, the USD/CHF exchange rate is an important information carrier during the 2007--2008 credit crisis. Over the same period, the AUD/JPY and NZD/JPY exchange rates change from being important for information transfer to being influential within their communities. The AUD/JPY and NZD/JPY were most influential within their community during Q3 and Q4 2007 and during Q1 and Q4 2008. Figure \ref{carry}(b) shows that over all of these periods there was significant carry trade activity so it is unsurprising that two exchange rates that are widely used for this trade should increase in importance. This, however, is a further demonstration that the positions of exchange rates in the $(z^b,z^y)$ parameter plane can provide important insights into the role of exchange rates in the FX market.

\section{Robustness of results: alternative computational heuristics}
\label{sec:heuristics}

Thus far, we have detected all communities using a greedy algorithm \cite{BLOND_ARX_2008}. However, as we noted in Section~\ref{sec::detection}, several alternative heuristics exist. We now investigate whether the choice of heuristic has any effect on the results described in this paper.

In Ref.~\cite{modcaution}, Good et al. demonstrated that there is extreme degeneracy in the energy function, with an exponential number of high-energy solutions. Given this, it is unsurprising that, in some instances, different energy-optimization heuristics have been found to yield very different partitions for the same network. Good et al. suggest that the reason for this behavior is that different heuristics sample different regions of the energy landscape. Because of the potential sensitivity of results to the choice of heuristic, one should treat individual partitions output by particular heuristics with caution. However, one can have more confidence in the validity of the partitions if different heuristics produce similar results.

In this section, we compare the results for the greedy algorithm \cite{BLOND_ARX_2008} with those for a spectral algorithm \cite{newmanpnas2006} and simulated annealing \cite{simulatedannealing} for the 563 networks we constructed for the period 2005--2008.

\subsection{Comparison of partition energies}

\begin{figure}
\centerline{\includegraphics[width=0.85\linewidth]{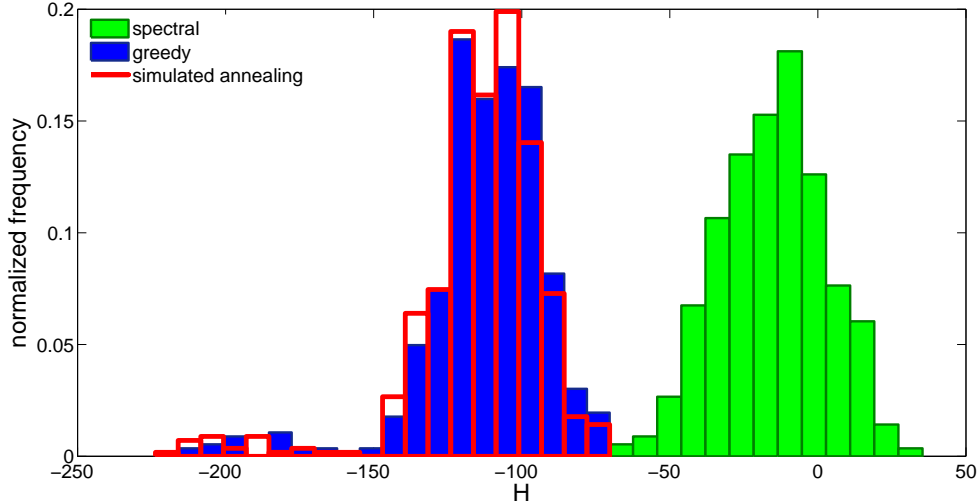}}
\caption[Distribution of $\mathcal{H}$ for different heuristics]{(Color online) Distribution of the energy $\mathcal{H}$ of the optimal partition for networks over the period 2005--2008 for different optimization algorithms.}\label{Hcomp_heuristics}
\end{figure}

We begin by comparing the energy $\mathcal{H}$ of the optimal partitions at the studied resolution $\gamma=1.45$. Figure~\ref{Hcomp_heuristics} shows the distribution of energies for the different algorithms and demonstrates that the greedy algorithm and simulated annealing find better partitions than the spectral algorithm. The spectral algorithm begins by splitting the network into two components, choosing the split that minimizes the energy, and then recursively partitions the smaller networks into two groups until no decrease in energy can be obtained through partitioning. At each step, the algorithm only finds the optimal partition of each community into two smaller communities, even though a split into more communities might yield a lower energy. Given this, it is unsurprising that the spectral algorithm identifies partitions further from the optimum than the other heuristics. For the remainder of this section, we will only compare the greedy and simulated annealing algorithms because of the lower quality of the spectral partitions.

\subsection{Temporal changes in communities}

\begin{figure}
\centerline{\includegraphics[width=0.85\linewidth]{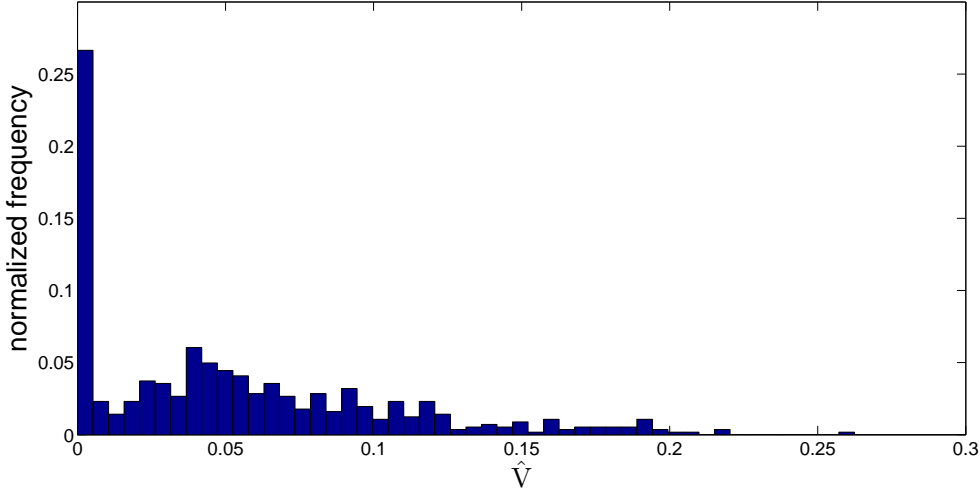}}
\caption[Distribution of variation of information between greedy and simulated annealing partitions]{(Color online) The distribution of the variation of information between community partitions identified using the greedy algorithm and simulated annealing for networks over the period 2005--2008.}\label{VI_heuristics}
\end{figure}

First, we compare the community partitions identified by the two heuristics for each network. In Fig.~\ref{VI_heuristics}, we show the distribution of the variation of information between the community partitions identified using the greedy and simulated annealing algorithms. The two methods identify identical partitions for 19\% of the networks; for 83\% of the networks, the partitions differ in their assignment of nodes to communities by fewer than 10 nodes. There is therefore strong agreement between the partitions obtained by the two heuristics, but there are also differences that warrant further investigation.

In Section~\ref{sec::majorchange}, we identified significant changes in the community configuration by comparing changes in the scaled energy $Q_s$ (see Eq.~\ref{Q_POTTS}) between consecutive time-steps and by calculating the variation of information between community partitions at consecutive time-steps (see Fig.~\ref{evolution}). The correlation between $Q_s$ as a function of time for the two heuristics is 0.99 and the correlation between the changes in $Q_s$ is 0.93. The correlation between the variation of information between partitions at consecutive time-steps is 0.36. The scaled energy correlations are clearly extremely high. However, there are differences in the timings of some major reorganizations identified by the variation of information. To compare the timings of major events, we identify time-steps at which the variation of information between consecutive partitions is more than a certain number of standard deviations larger the mean variation of information between consecutive partitions. We find that the algorithms identify 40\% of one standard deviation events at the same time-steps and 33\% of 2.5 standard deviation events. The methods therefore agree reasonably well, with one in three 2.5 standard deviation events identified at exactly the same time-step. However, the differences also suggest that one should be cautious using variation of information to identify major community reorganizations.

\subsection{Example community comparison}

\begin{figure}[tbp]
\begin{center}
\centerline{\includegraphics[width=0.85\linewidth]{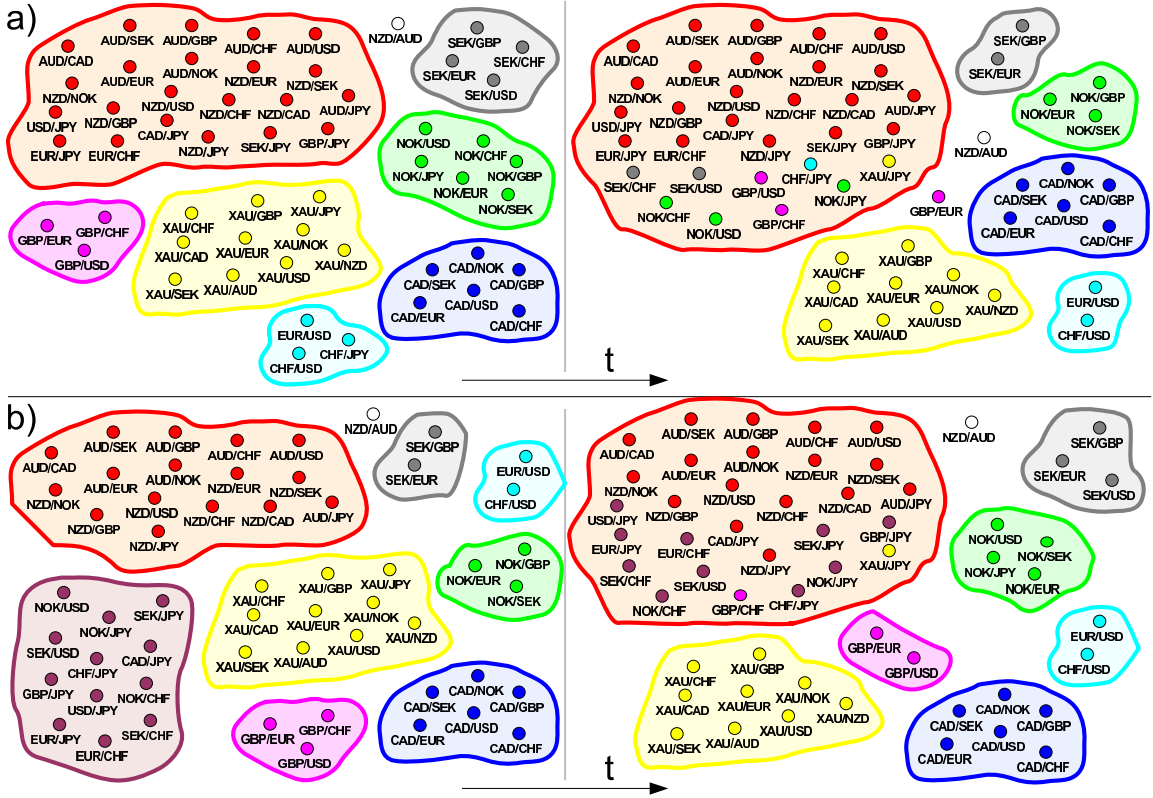}}
\caption[Heuristic comparison: community change schematic]{\label{comm_schematic_algo}(Color online) Comparison of the change in community structure in one half of the FX market network over the same period for different optimization heuristics. We show a schematic of the communities for the period following 08/15/07, when there was significant unwinding of the carry trade during the 2007--2008 credit and liquidity crisis. We identified communities using (a) a greedy algorithm \cite{BLOND_ARX_2008} and (b) a simulated annealing algorithm \cite{simulatedannealing}. The node colors after the community reorganization correspond to their community before the change. If the parent community of a community after the reorganization is obvious, we draw it using the same color as its parent. The nodes drawn as triangles resided in the opposite half of the network before the community reorganization.}
\end{center}
\end{figure}

One time-step at which both heuristics identify a large community change is 08/15/07 which, as described in Section~\ref{subsec:carry}, was a day when there was a significant increase in carry trade unwinding. It is worth considering the communities at this time-step in detail to help assess the similarity of the results for the two heuristics. In Fig.~\ref{comm_schematic_algo}(a), we show the communities that we identified using a greedy algorithm \cite{BLOND_ARX_2008} immediately before and after 08/15/07; in Fig.~\ref{comm_schematic_algo}(b) we show communities that we identified using simulated annealing \cite{simulatedannealing} for the same time-steps.  Figure~\ref{comm_schematic_algo}(a) shows that, leading up to 08/15/07, there was some unwinding of the carry trade, so the initial configuration includes a community containing exchange rates of the form AUD/YYY, NZD/YYY, and XXX/JPY (which all involve one of the key carry-trade currencies). After 08/15/07, as the volume of carry trade unwinding increases, this community incorporates other XXX/JPY rates as well as some XXX/CHF and XXX/USD rates. Although, the communities in Fig.~\ref{comm_schematic_algo}(b) for the simulated annealing algorithm are not identical to those in Fig.~\ref{comm_schematic_algo}(a), they are very similar. The main difference is that for the simulated annealing algorithm, there are two carry trade communities before 08/15/07: one community of exchange rates of the form AUD/YYY, NZD/YYY (which are all exchange rates that include a carry trade investment currency) and another community containing exchange rates of the form XXX/CHF and XXX/JPY (which are all exchange rates that include a carry trade funding currency). After 08/15/07, as carry trade unwinding increases, these two communities combine and two other exchange rates also join the community. The resulting merged community is very similar to the largest community identified at the same time-step using the greedy algorithm.

Figure~\ref{comm_schematic_algo} therefore illustrates that there are only small differences in the community configurations that are identified by the two heuristics. In fact, as Fig.~\ref{VI_heuristics} shows, the two algorithms agree in the assignment of all but about ten nodes approximately 80\% of the time. Importantly, Fig.~\ref{comm_schematic_algo} highlights that, even when there are differences in the exact community configurations, the communities that are identified by the two heuristics nonetheless indicate the same changes taking place in the FX market.

\subsection{Node role comparison}

\begin{figure}
\centerline{\includegraphics[width=0.85\linewidth]{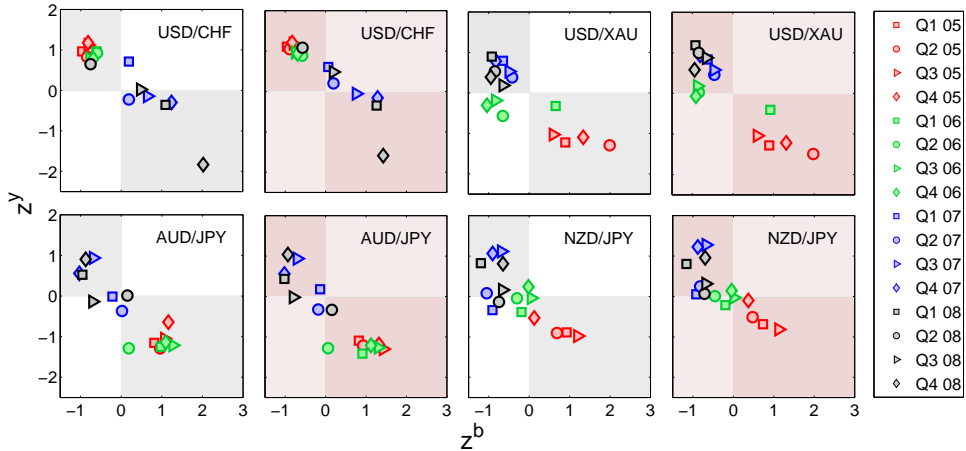}}
\caption[Heuristic comparison: quarterly node roles]{(Color online) Comparison of the quarterly node role evolutions in the $(z^b,z^y)$ plane for the period 2005--2008 for communities identified using a greedy algorithm \cite{BLOND_ARX_2008} and simulated annealing \cite{simulatedannealing}. The white/gray shading plots show results for the greedy algorithm and the pink/dark pink plots show results for simulated annealing.}\label{heuristicroles}
\end{figure}

As a further comparison, we investigate the effect of different heuristics on exchange rate roles (see Section~\ref{sec::rolevis}). In Fig.~\ref{heuristicroles}, we compare quarterly role evolutions over the period 2005--2008 for the same exchanges rate shown in Fig.~\ref{quarterlyroles}. Although there are slight differences in the positions of the exchange rates in the $(z^b,z^y)$ plane for some periods, we obtain the same aggregate conclusions. For example, for both heuristics, AUD/JPY is most influential within its community (high $z^b$) during Q3 and Q4 2007 and during Q1 and Q4 2008 and is less influential, but more important for information transfer, during 2005 and 2006.

The positions in the $(z^b,z^y)$ plane are similarly close for all of the other exchange rates. We quantify the differences in the positions for the two heuristics by calculating the mean and standard deviation of the change in position over all exchange rates and over all time periods. That is, we average the change in position of every node in the $(z^b,z^y)$ plane over every quarter. The mean change in position in both the $z^b$ and $z^y$ directions is less than $10^{-4}$; the standard deviations are 0.15 and 0.17, respectively. However, because the changes in position are likely to cancel out (i.e., an increase in $z^b$ for one exchange rate is likely to be offset by a decrease in $z^b$ for another exchange rate), it is more informative to calculate the mean and standard deviation of the absolute changes in position in the $z^b$ and $z^y$ directions. In the $z^b$ direction, the mean absolute change in position is 0.08, with a standard deviation of 0.13; in the $z^y$ direction, the mean change is 0.09, with a standard deviation of 0.15. The mean differences in positions in the $(z^b,z^y)$ are therefore very small for the two heuristics and, as Fig.~\ref{heuristicroles} demonstrates, both algorithms uncover the same role changes in the FX market for the different exchange rates.

Finally, we also checked the relationships shown in Fig.~\ref{station} between the community centrality and community size, between the community alignment and betweenness centrality, and between the community autocorrelation and projected community centrality. Using simulated annealing, we find the same relationships that we uncovered with the greedy algorithm.

The results of this section demonstrate that, although there are differences in the communities identified using different optimization heuristics, the aggregate conclusions are the same. We identify the same changes taking place in the FX market whether we use the greedy algorithm or simulated annealing to minimize energy. The fact that we obtain very similar results using different optimization techniques, despite these techniques sampling different regions of the energy landscape, gives confidence that the effects that we uncover are genuine and that our results are robust. In practice, the greedy algorithm is preferable to simulated annealing because of the computational cost of the latter. For example, the greedy algorithm converges on an optimal community partition for all 563 networks over the period 2005--2008 in 5 minutes 24 seconds. For the same networks, the simulated annealing algorithm takes 36 hours.

%%%%%%%%%%%%%%%%%%%%%%%%%%%%%%%%%%%%%%%%%%%%

\section{Conclusions} \label{sec::conclusions}

To conclude, we have demonstrated that a network analysis of the FX market is useful for visualizing and providing insights into the correlation structure of the market. In particular, we investigated community structure at different times to provide insights into the clustering dynamics of exchange rate time series. We focused on a node-centric community analysis that allows one to follow the time dynamics of the functional role of exchange rates within the market, demonstrating that there is a relationship between an exchange rate's functional role and its position within its community. We indicated that exchange rates that are located on the edges of communities are important for information transfer in the FX market, whereas exchange rates that are located in the center of their community have a strong influence on other rates within that community. We also demonstrated that the community structure of the market can be used to determine which exchange rates dominate the market at each time-step and identified exchange rates that experienced significant changes in market role.

Our analysis successfully uncovered significant structural changes that occurred in the FX market, including ones that resulted from major market events that did not impact the studied exchange rates directly. We further demonstrated that community reorganizations at specific time-steps can provide insights into changes in trading behavior and highlighted the prevalence of the carry trade during the 2007--2008 credit and liquidity crisis. Although we focused on networks of exchange rates, our methodology should be similarly insightful for multivariate time series of other asset classes.

%%%%%%%%%%%%%%%%%%%%%%%%%%%%%%%%%%%%%%%%%%%%

\section*{Acknowledgments}

We thank M. Gould, S.~D. Howison, A.~C.~F. Lewis, M.~A. Little, J.~M. McPherson, J. Moody, J.-P. Onnela, S. Reid, and S.~J. Roberts and his group for discussions and code. We acknowledge HSBC bank for providing the data. N.~S.~J. acknowledges the BBSRC and EPSRC. M.~A.~P. acknowledges a research award (\#220020177) from the James S. McDonnell Foundation. P.~J.~M.'s contribution was funded by the NSF (DMS-0645369).

%%%%%%%%%%%%%%%%%%%%%%%%%%%%%%%%%%%%%%%%%%%%

%\bibliographystyle{harvard}
\bibliography{Fenn_et_al_FXComm}

\end{document}